\documentclass[aps,prb,twocolumn,superscriptaddress,floatfix]{revtex4}%
\bibpunct{[}{]}{,}{n}{}{}
\usepackage{graphicx}
\usepackage{dcolumn}
\usepackage{latexsym}
\usepackage{lipsum}
\usepackage{amssymb}
\usepackage{amsfonts}
\usepackage{amsmath}
\usepackage{fancybox}
\usepackage[binary,squaren]{SIunits}
\usepackage{color}
\usepackage{colordvi}%
\providecommand{\U}[1]{\protect\rule{.1in}{.1in}}

\def\be{\begin{equation}}
\def\ee{\end{equation}}
\def\ber{\begin{eqnarray}}
\def\eer{\end{eqnarray}}

\begin{document}
\title{Magnon-polaron transport in magnetic insulators}
\author{Benedetta Flebus}
\affiliation{Institute for Theoretical Physics and Center for Extreme Matter and Emergent
Phenomena, Utrecht University, Leuvenlaan 4, 3584 CE Utrecht, The Netherlands}
\author{Ka Shen}
\affiliation{Kavli Institute of NanoScience, Delft University of Technology, Lorentzweg 1,
2628 CJ Delft, The Netherlands}
\author{Takashi Kikkawa}
\affiliation{Institute for Materials Research, Tohoku University, Sendai 980-8577, Japan}
\affiliation{WPI Advanced Institute for Materials Research, Tohoku University, Sendai
980-8577, Japan}
\author{Ken-ichi Uchida}
\affiliation{Institute for Materials Research, Tohoku University, Sendai 980-8577, Japan}
\affiliation{National Institute for Materials Science, Tsukuba 305-0047, Japan}
\affiliation{PRESTO, Japan Science and Technology Agency, Saitama 332-0012, Japan}
\affiliation{Center for Spintronics Research Network, Tohoku University, Sendai 980-8577, Japan}
\author{Zhiyong Qiu}
\affiliation{WPI Advanced Institute for Materials Research, Tohoku University, Sendai
980-8577, Japan}
\author{Eiji Saitoh}
\affiliation{Institute for Materials Research, Tohoku University, Sendai 980-8577, Japan}
\affiliation{WPI Advanced Institute for Materials Research, Tohoku University, Sendai
980-8577, Japan}
\affiliation{Center for Spintronics Research Network, Tohoku University, Sendai 980-8577, Japan}
\affiliation{Advanced Science Research Center, Japan Atomic Energy Agency, Tokai 319-1195, Japan}
\author{Rembert A. Duine}
\affiliation{Institute for Theoretical Physics and Center for Extreme Matter and Emergent
Phenomena, Utrecht University, Leuvenlaan 4, 3584 CE Utrecht, The Netherlands}
\affiliation{Department of Applied Physics, Eindhoven University of Technology, PO Box 513,
5600 MB Eindhoven, The Netherlands}
\author{Gerrit E. W. Bauer}
\affiliation{Institute for Materials Research, Tohoku University, Sendai 980-8577, Japan}
\affiliation{WPI Advanced Institute for Materials Research, Tohoku University, Sendai
980-8577, Japan}
\affiliation{Kavli Institute of NanoScience, Delft University of Technology, Lorentzweg 1,
2628 CJ Delft, The Netherlands}
\affiliation{Center for Spintronics Research Network, Tohoku University, Sendai 980-8577, Japan}
\date{\today}

\begin{abstract}
We theoretically study the effects of strong magnetoelastic coupling on the
transport properties of magnetic insulators. We develop a Boltzmann transport
theory for the mixed magnon-phonon modes ("magnon polarons") and determine
transport coefficients and spin diffusion length. Magnon-polaron formation
causes anomalous features in the magnetic field and temperature dependence of the spin
Seebeck effect when the disorder scattering in the magnetic and elastic subsystems is sufficiently
different. Experimental data by Kikkawa \textit{et al.} [PRL
\textbf{117}, 207203 (2016)] on yttrium iron garnet 
films can be explained by an acoustic quality that is much better than the
magnetic quality of the material. We predict similar anomalous features in the
spin and heat conductivity and non-local spin transport experiments.

\end{abstract}
\maketitle


\section{Introduction}

The magnetoelastic coupling (MEC) between magnetic moments and lattice
vibrations in ferromagnets stems from spin-orbit, dipole-dipole and exchange
interactions. This coupling gives rise to magnon-polarons, i.e., hybridized
magnon and phonon modes in proximity of the intersection of the uncoupled
elastic and magnetic dispersions~\cite{Kittelfirst,MECref, Kittel58,
kalga}. Interest in the coupling of magnetic and elastic excitations emerged
recently in the field of spin caloritronics~\cite{spinc}, since it affects thermal and
spin transport properties of magnetic insulators such as yttrium iron garnet
(YIG)~\cite{hillebrandsexperiment,spinpumpingacoustic,spinpumpingacoustic1,akash,Ogawa,shen16}. 

In this work we address the spin Seebeck effect (SSE) at low temperatures -- which
provides an especially striking evidence for magnon-polarons in the form of
asymmetric spikes in the magnetic field dependence~\cite{kikkawa}. The
enhancement emerges at the magnetic fields corresponding to the tangential
intersection of the magnonic dispersion with the acoustic longitudinal and
transverse phonon branches that we explain by phase-space arguments and an
unexpected high acoustic quality of YIG.

Here we present a Boltzmann transport theory for coupled magnon and phonon
transport in bulk magnetic insulators and elucidate the anomalous field and
temperature dependencies of the SSE in terms of the composite nature of the
magnon-polarons. The good agreement between theory and the experiments
generates confidence that the SSE can be used as an instrument to characterize
magnons vs. phonon scattering in a given material. We derive the full Onsager
matrix, including spin and heat conductivity as well as the spin diffusion
length. We predict magnon-polaron signatures in all transport coefficients
that await experimental exposure.

This work is organized as follows: In Sec.~II we start by introducing the
standard model for spin wave and phonon band dispersions of a magnetic
insulator and the magnetoelastic coupling. In Sec.~III, we describe the
magnon-polaron modes and their field-dependent behavior in reciprocal space.
The linearized Boltzmann equation is shown to lead to expressions for the
magnon-polaron transport coefficients. In Sec.~IV, we present numerical
results for the spin Seebeck coefficient, spin and heat conductivity, and spin
diffusion length for YIG. We also derive approximate analytical expressions
for the field and temperature dependence of the anomalies emerging in the
transport coefficients and compare our results with the experiments. In Sec.~V
we present our conclusions and an outlook.


\section{Model}

In this section we introduce the Hamiltonian describing the coupling between magnons and phonons in 
 magnetic insulators. The experimentally relevant geometry is  schematically depicted in Fig.~\ref{scheme}.

\subsection{Magnetic Hamiltonian}

\begin{figure}[ptb]
\includegraphics[width=6.5cm]{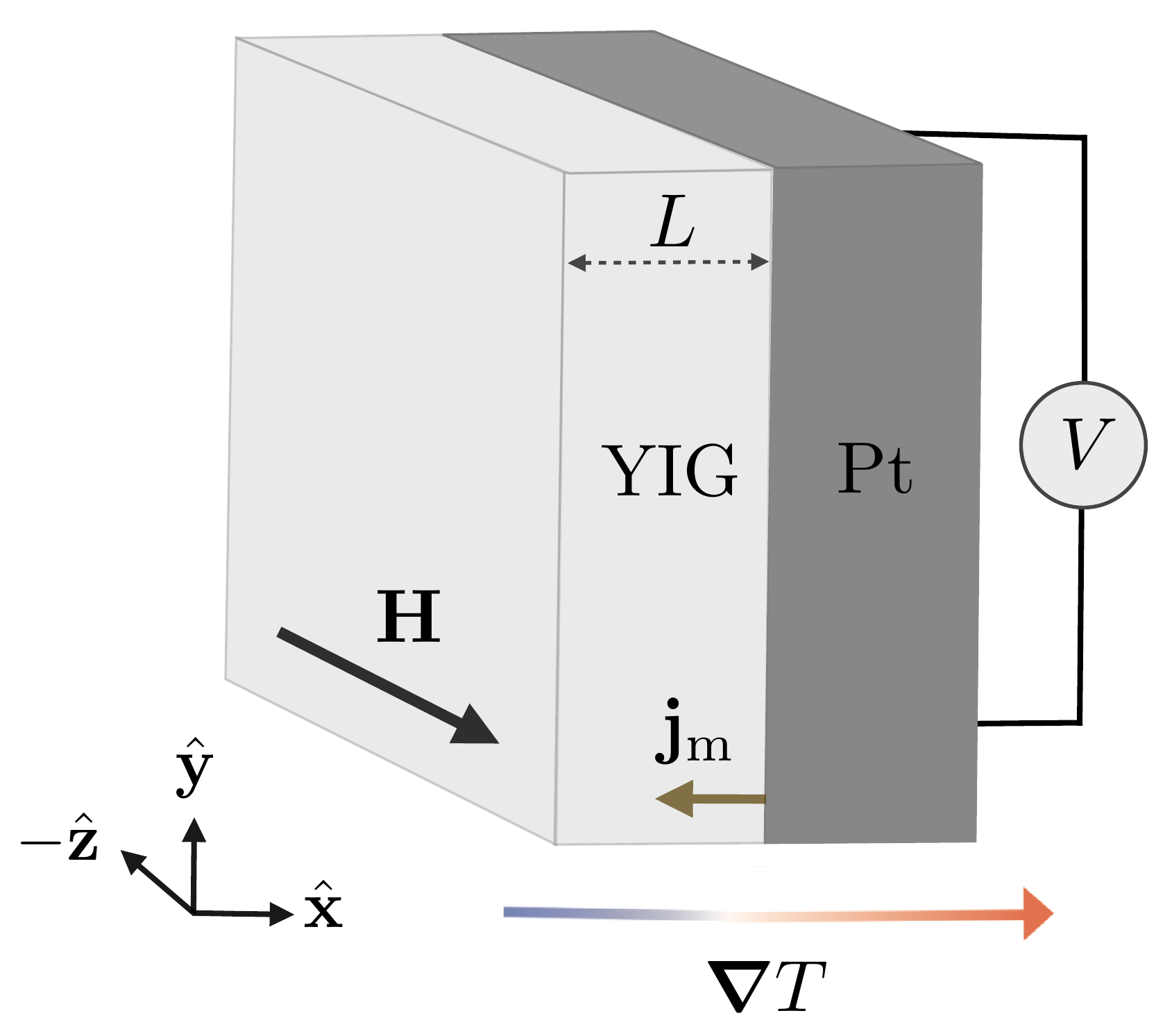} 
\caption{Pt$|$YIG bilayer subject to a thermal gradient $\boldsymbol{\nabla} T \parallel \hat{\mathbf{x}}$ and a magnetic field $ \mathbf{H} \parallel \hat{\mathbf{z}}$. The thermal bias gives rise to a flow of magnons, i.e., a magnonic spin current $\mathbf{j}_{\text{m}}$, in the YIG film of thickness $L$. In the Pt lead, the spin current is then converted into a measurable voltage $V$ via the inverse Spin Hall effect. }
\label{scheme}%
\end{figure}

We consider a magnetic insulator with spins $\mathbf{S}_{p}=\mathbf{S}%
(\mathbf{r}_{p})$ localized on lattice sites $\mathbf{r}_{p}$. The magnetic
Hamiltonian consists of dipolar and (Heisenberg) exchange interactions between
spins and of the Zeeman interaction due to an external magnetic field $\mathbf{B}%
=\mu_{0}H\mathbf{\hat{z}}$~\cite{Akhiezer,ham, ham1}. 
It reads as
\begin{align}
\mathcal{H}_{\mathrm{mag}}\hspace*{-0.1cm} &  =\frac{\mu_{0}(g\mu_{B})^{2}}%
{2}\sum_{p\neq q}\frac{|\mathbf{r}_{pq}|^{2}{\mathbf{S}_{p}\cdot\mathbf{S}%
_{q}}-3\left(  \mathbf{r}_{pq}\cdot\mathbf{S}_{p}\right)  \left(
\mathbf{r}_{pq}\cdot\mathbf{S}_{q}\right)  }{|\mathbf{r}_{pq}|^{5}}\nonumber\\
&  -J\sum_{p\neq q}\mathbf{S}_{p}\cdot\mathbf{S}_{q}\,-g\mu_{B}B\sum_{p}%
S_{p}^{z}\,.\label{spinH}%
\end{align}
Here, $g$ is the g-factor, $\mu_{0}$ the vacuum permeability, $\mu_{B}$ the
Bohr magneton, $J$ the exchange interaction strength, and $\mathbf{r}_{pq}%
=\mathbf{r}_{p}-\mathbf{r}_{q}$. By averaging over the complex unit cell of a
material such as YIG, we define a coarse-grained, classical spin $S=\left\vert
\mathbf{S}_{p}\right\vert =a_{0}^{3}M_{s}/(g\mu_{B})$ on a cubic lattice with
unit cell lattice constant $a_{0}$, with $M_{s}$ being the zero temperature saturation
magnetization density. The crystal anisotropy is disregarded, while the
dipolar interaction is evaluated for a  magnetic film in the $yz$-plane, see Fig.~\ref{scheme}.
 We employ the Holstein-Primakoff transformation and expand the
spin operators as~\cite{holstein40}
\begin{align}
&  S_{p}^{-}=\sqrt{2S}a_{p}^{\dagger}\sqrt{1-\frac{a_{p}^{\dagger}a_{p}}{2S}%
}\approx\sqrt{2S}\left[  a_{p}^{\dagger}-\frac{a_{p}^{\dagger}a_{p}^{\dagger
}a_{p}}{4S}\right]\,,\nonumber\\
&  S_{p}^{z}=S-a_{p}^{\dagger}a_{p}\,,\label{HP}%
\end{align}
where $S_{p}^{-}=S_p^x -iS_p^y$, and
$a_{p}$/$a_{p}^{\dagger}$ annihilate/create a magnon at the lattice site $\mathbf{r}_{p}$ and obey Boson
commutation rules $[a_{p},a_{q}^{\dagger}]=\delta_{pq}$. Substituting the
Fourier representation
\begin{equation}
a_{p}=\frac{1}{\sqrt{N}}\sum_{\mathbf{k}}e^{i\mathbf{k}\cdot\mathbf{r}_{p}%
}a_{\mathbf{k}}\,,\;\;\;a_{p}^{\dagger}=\frac{1}{\sqrt{N}}\sum_{\mathbf{k}%
}e^{-i\mathbf{k}\cdot\mathbf{r}_{p}}a_{\mathbf{k}}^{\dag}\,,
\end{equation}
where $N$ is the number of lattice sites, and retaining only quadratic terms in
the bosonic operators and disregarding a constant, the
Hamiltonian~(\ref{spinH}) becomes
\begin{equation}
\mathcal{H}_{\mathrm{mag}}=\sum_{{\mathbf{k}}}A_{\mathbf{k}}a_{{\mathbf{k}}%
}^{\dagger}a_{{\mathbf{k}}}+\frac{1}{2}\left(B_{\mathbf{k}}a_{-\mathbf{k}}a_{\mathbf{k}}+B_{\mathbf{k}}^\ast  a_{\mathbf{k}%
}^{\dagger}a_{-\mathbf{k}}^{\dagger}\right)\,,\label{spinH2}%
\end{equation}
with
\begin{align}
\frac{A_{\mathbf{k}}}{\hbar} &  =D_{\mathrm{ex}}\mathcal{F}_{\mathbf{k}%
}+\gamma\mu_{0}H+\frac{\gamma\mu_{0}M_{s}\sin^{2}\theta_{\mathbf{k}}}%
{2}\,,\nonumber\\
\frac{B_{\mathbf{k}}}{\hbar} &  =\frac{\gamma\mu_{0}M_{s}\sin^{2}%
\theta_{\mathbf{k}}}{2} e^{-2 i \phi_{\mathbf{k}}}\,.
\end{align}
Here, $D_{\mathrm{ex}}=2SJa_{0}^{2}$ is the exchange stiffness, $\gamma
=g\mu_{B}/\hbar$ the gyromagnetic ratio, $\theta_{\mathbf{k}}=\arccos\left(
k_{z}/k\right)  $ 
the polar angle between wave-vector $\mathbf{k}$ with
$k=|\mathbf{k}|$ and the magnetic field along $\mathbf{\hat{z}}$ and $\phi_{\mathbf{k}}$  the azimuthal angle of $\mathbf{k}$ in the $xy$
plane. The form
factor $\mathcal{F}(\mathbf{k})=2(3-\cos k_{x}a_{0}-\cos k_{y}a_{0}-\cos
k_{z}a_{0})/a_{0}^{2}$ can be approximated as $\mathcal{F}(\mathbf{k})\approx
k^{2}$ in the long-wavelength limit ($ka_{0}\ll1$). Equation~(\ref{spinH2}) is
diagonalized by the Bogoliubov transformation~\cite{Keffer66}
\begin{equation}%
\begin{bmatrix}
a_{\mathbf{k}}\\
a_{-\mathbf{k}}^{\dagger}%
\end{bmatrix}
=%
\begin{bmatrix}
u_{\mathbf{k}} & -v_{\mathbf{k}}\\
-v_{\mathbf{k}}^{\ast} & u_{\mathbf{k}}%
\end{bmatrix}%
\begin{bmatrix}
\alpha_{\mathbf{k}}\\
\alpha_{-\mathbf{k}}^{\dagger}%
\end{bmatrix}\,,
\label{BG}%
\end{equation}
with parameters
\begin{equation}
u_{\mathbf{k}}=\sqrt{\frac{A_{\mathbf{k}}+\hbar\omega_{\mathbf{k}}}%
{2\hbar\omega_{\mathbf{k}}}}\,,\;\;\;v_{\mathbf{k}}=\sqrt{\frac{A_{\mathbf{k}%
}-\hbar\omega_{\mathbf{k}}}{2\hbar\omega_{\mathbf{k}}}}e^{2i\phi_{\mathbf{k}}%
}\,.\label{BGcoeff}%
\end{equation}
The Hamiltonian (\ref{spinH2}) is then simplified to
\begin{equation}
\mathcal{H}_{\mathrm{mag}}=\sum_{{\mathbf{k}}}\hbar\omega_{\mathbf{k}}%
\alpha_{{\mathbf{k}}}^{\dagger}\alpha_{{\mathbf{k}}}\,,
\end{equation}
where $\hbar\omega_{\mathbf{k}}=\sqrt{A_{\mathbf{k}}^{2}-|B_{\mathbf{k}}|^{2}}$ 
is the magnon dispersion. 
For bulk magnons in the long-wavelength limit~\cite{Schlomann60,StaticsPhysics}
\begin{equation}
\omega_{\mathbf{k}}=\sqrt{D_{\mathrm{ex}}k^{2}+\gamma\mu_{0}H}\sqrt
{D_{\mathrm{ex}}k^{2}+\gamma\mu_{0}(H+M_{s}\sin^{2}\theta_{\mathbf{k}}%
)}.\label{magnond}%
\end{equation}
We disregard Damon-Eshbach modes~\cite{Damon1961} localized at the surface since, in the following,
we focus  on transport in thick films normal to the plane, i.e., in the \textit{x}-direction in Fig.~\ref{scheme}. For
thick films the backward moving volume modes are relevant only for wave
numbers $k$ very close to the origin and are disregarded as well. Higher order
terms in the magnon operators that encode magnon-magnon scattering processes
have been disregarded as well in Eq.~(\ref{spinH2}), which is allowed for
sufficiently low magnon-densities or temperatures (for YIG $\lesssim100\,$K~\cite{Barker2016}). 
In this regime, the main relaxation mechanism is magnon scattering by static disorder~\cite{hillebrandsexperiment} with Hamiltonian
\begin{equation}
\mathcal{H}_{\mathrm{mag{\text -}imp}}=\sum_{\mathbf{k},\mathbf{k}^{\prime}%
}v_{\mathbf{k},\mathbf{k}^{\prime}}^{\mathrm{mag}}\alpha_{\mathbf{k}}^{\dagger
}\alpha_{\mathbf{k}^{\prime}}\,,\label{Himpurity}%
\end{equation}
where $v_{\mathbf{k},\mathbf{k}^{\prime}}^{\mathrm{mag}}$ is an
impurity-scattering potential. In the following, we employ the isotropic,
short-range scattering approximation $v_{\mathbf{k},\mathbf{k}^{\prime}%
}^{\mathrm{mag}}=v^{\mathrm{mag}}$.

\subsection{Mechanical Hamiltonian}

We focus on lattice vibrations or sound waves with wavelengths much larger
than the lattice constant that are well-described by continuum mechanics. The
Hamiltonian of an elastically isotropic solid reads~\cite{Kittel63}
\begin{align}
\mathcal{H}_{\mathrm{el}}  &  =\int d^{3}r\sum_{i,j}\frac{\Pi_{i}%
^{2}(\mathbf{r})}{2 \bar{\rho}}\delta_{ij}+(c_{\parallel}^{2}-c_{\bot}^{2})\frac
{\bar{\rho}}{2}\frac{\partial R_{i}(\mathbf{r})}{\partial x_{i}}\frac{\partial
R_{j}(\mathbf{r})}{\partial x_{j}}\nonumber\\
&  +c_{\parallel}^{2}\frac{\bar{\rho}}{2}\frac{\partial R_{i}(\mathbf{r}%
)}{\partial x_{j}}\frac{\partial R_{i}(\mathbf{r})}{\partial x_{j}},
\label{13}%
\end{align}
where $\bar{\rho}$ is the average mass density, $R_{i}$ is the $i$-th
component  of the displacement vector
$\mathbf{R}$ of a volume element at $\mathbf{r}$ with respect to its
equilibrium position, $\Pi_{i}$ is the conjugate phonon momentum and
$c_{\parallel}$ and $c_{\bot}$ are the velocities of the
longitudinal acoustic (LA) and transverse acoustic (TA) lattice waves, respectively. The
Hamiltonian (\ref{13}) can be quantized by the phonon creation (annihilation)
operators $c_{\lambda{\mathbf{k}}}^{\dagger}$ ($c_{\lambda{\mathbf{k}}}$) as
\begin{align}
R_{i}(\mathbf{r},t)  &  =\sum_{{\mathbf{k}},\lambda}\epsilon_{i\lambda}({\mathbf{k}%
})\left(  \frac{\hbar}{2\bar{\rho}V\omega_{\lambda\mathbf{k}}}\right)
^{1/2}(c_{\lambda{\mathbf{k}}}^{\dagger}+c_{\lambda{-\mathbf{k}}})e^{i{\mathbf{k}%
}\mathbf{r}}\,,\\
\Pi_{i}(\mathbf{r},t)  &  =i\sum_{{\mathbf{k}},\lambda}\epsilon_{i\lambda}%
({\mathbf{k}})\left(  \frac{\bar{\rho}\hbar\omega_{\lambda\mathbf{k}}}{2V}\right)
^{1/2}\left(  c_{\lambda{\mathbf{k}}}^{\dagger}-c_{\lambda-{\mathbf{k}}}\right)
e^{-i{\mathbf{k}}\mathbf{r}}\,, \label{Pi}%
\end{align}
where $\lambda=1,2$ labels the shear waves polarized normal to the wave-vector
${\mathbf{k}}$ (TA phonons), while $\lambda=3$ represents a pressure wave (LA
phonons). Here $\omega_{\lambda\mathbf{k}}=c_{\lambda}|\mathbf{k}|$ is the phonon
dispersion and $\epsilon_{i\lambda}(\mathbf{k})=\hat{\mathbf{x}}_{i}\cdot
\hat{\epsilon}(\mathbf{k},\lambda)$ are Cartesian components $i=x,y,z$ of the unit
polarization vectors
\begin{subequations}
\begin{align}
\hat{\epsilon}(\mathbf{k},1)  &  =(\cos\theta_{\mathbf{k}}\cos\phi
_{\mathbf{k}},\cos\theta_{\mathbf{k}}\sin\phi_{\mathbf{k}},-\sin
\theta_{\mathbf{k}})\,,\\
\hat{\epsilon}(\mathbf{k},2)  &  =i(-\sin\phi_{\mathbf{k}},\cos\phi
_{\mathbf{k}},0)\,,\\
\hat{\epsilon}(\mathbf{k},3)  &  =i(\sin\theta_{\mathbf{k}}\cos\phi
_{\mathbf{k}},\sin\theta_{\mathbf{k}}\sin\phi_{\mathbf{k}},\cos\theta
_{\mathbf{k}})\,,
\end{align}
that satisfy $\hat{\epsilon}^{\ast}(\mathbf{k},\lambda)=\hat{\epsilon}%
(-\mathbf{k},\lambda)$~\cite{hillebrandsexperiment}. In terms of the operators
$c_{\lambda{\mathbf{k}}}$ and $c_{\lambda{\mathbf{k}}}^{\dagger}$, Eq.~(\ref{13})
becomes
\end{subequations}
\begin{equation}
\mathcal{H}_{\mathrm{el}}=\sum_{\mathbf{k},\lambda}\hbar\omega_{\lambda\mathbf{k}%
}\left(  c_{\lambda{\mathbf{k}}}^{\dagger}c_{\lambda{\mathbf{k}}}+\tfrac{1}{2}\right)
\,.
\end{equation}

Analogous to magnons, at low temperatures phonon relaxation is dominated by
static disorder
\begin{equation}
\mathcal{H}_{\mathrm{imp}}=\sum_{\lambda}\sum_{\mathbf{k},\mathbf{k}^{\prime}%
}v_{\mathbf{k},\mathbf{k}^{\prime}}^{\mathrm{ph}}c_{\lambda\mathbf{k}}^{\dagger
}c_{\lambda
\mathbf{k}^{\prime}}\,, \label{Himpurityphonon}%
\end{equation}
where $v_{\mathbf{k},\mathbf{k}^{\prime}}^{\mathrm{ph}}$ is the phonon
impurity-scattering potential, in the following assumed to be isotropic and
short-range, i.e., $v_{\mathbf{k},\mathbf{k}^{\prime}}^{\mathrm{ph}%
}=v^{\mathrm{ph}}$.

\subsection{Magnetoelastic coupling}

The magnetic excitations are coupled to the elastic displacement via
magnetoelastic interactions. In the long-wavelength limit, to leading order in
the magnetization $M_{i}=ng\mu_{B}S_{i}$ ($n=1/a_{0}^{3}$) and displacement
field $R_{i}$, the magnetoelastic energy reads as~\cite{Kittel58,Keffer66}
\begin{align}
\mathcal{H}_{\mathrm{mec}}  &  =\frac{\hbar n}{M_{s}^{2}}\int d^{3}r\sum
_{ij}\left[  B_{ij}M_{i}(\mathbf{r})M_{j}(\mathbf{r})\right. \nonumber\\
&  \left.  +B_{ij}^{^{\prime}}\frac{\partial\mathbf{M}(\mathbf{r})}{\partial
r_{i}}\cdot\frac{\partial\mathbf{M}(\mathbf{r})}{\partial r_{j}}\right]
R_{ij}(\mathbf{r})\,, \label{MEC}%
\end{align}
where $B_{ij}=\delta_{ij}B_{\parallel}+(1-\delta_{ij})B_{\bot}$ and
$B_{ij}^{\prime}=\delta_{ij}B_{\parallel}^{\prime}+(1-\delta_{ij})B_{\bot
}^{\prime}$ are the phenomenological magnetoelastic constants and
\begin{equation}
R_{ij}(\mathbf{\mathbf{r}})=\frac{1}{2}\left[  \frac{\partial R_{i}%
(\mathbf{r})}{\partial r_{j}}+\frac{\partial R_{j}(\mathbf{r})}{\partial
r_{i}}\right]\,,
\end{equation}
is the displacement gradient $R_{ij}$.

The exchange term $\sim B_{ij}^{\prime}$ in Eq.~(\ref{MEC}) contains
magnetization gradients and  predominantly affects short wavelength magnons. We
disregard this term since we are interested in capturing low temperature
features. Linearizing with respect to small nonequilibrium variables --
$R_{i}$, $M_{x},M_{y}$ -- Eq.~(\ref{MEC}) then becomes
\begin{align}
\mathcal{H}_{\mathrm{mec}}  &  =\hbar nB_{\perp}\left(  \frac{\gamma\hbar^{2}
}{4M_{s}\bar{\rho}}\right)  ^{1/2}\sum_{{\mathbf{k}},\lambda}k\omega
_{\mathbf{k}\lambda}^{-1/2}e^{-i\phi}a_{\mathbf{k}}(c_{\lambda-\mathbf{k}}%
+c_{\lambda\mathbf{k}}^{\dagger})\nonumber\\
&  \times(-i\delta_{\lambda1}\cos2\theta_{\mathbf{k}}+i\delta_{\lambda2}\cos
\theta_{\mathbf{k}}-\delta_{\lambda3}\sin2\theta_{\mathbf{k}})+\mathrm{H.c.}\,,
\label{hamiltME}%
\end{align}
where $\delta_{\lambda i}$ is the Kronecker delta.

\begin{figure}[ptb]
\includegraphics[width=6cm]{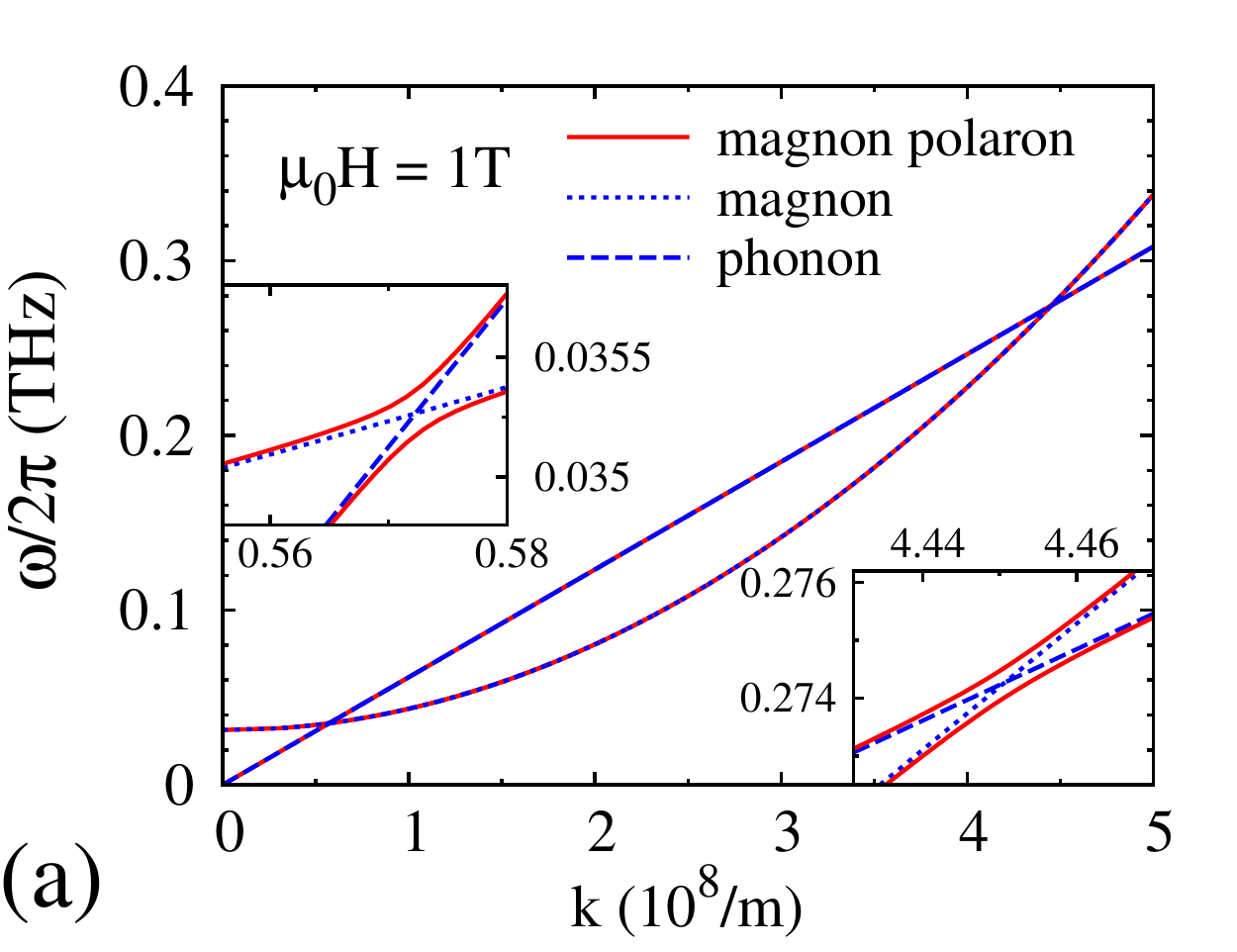} 
\includegraphics[width=6cm]{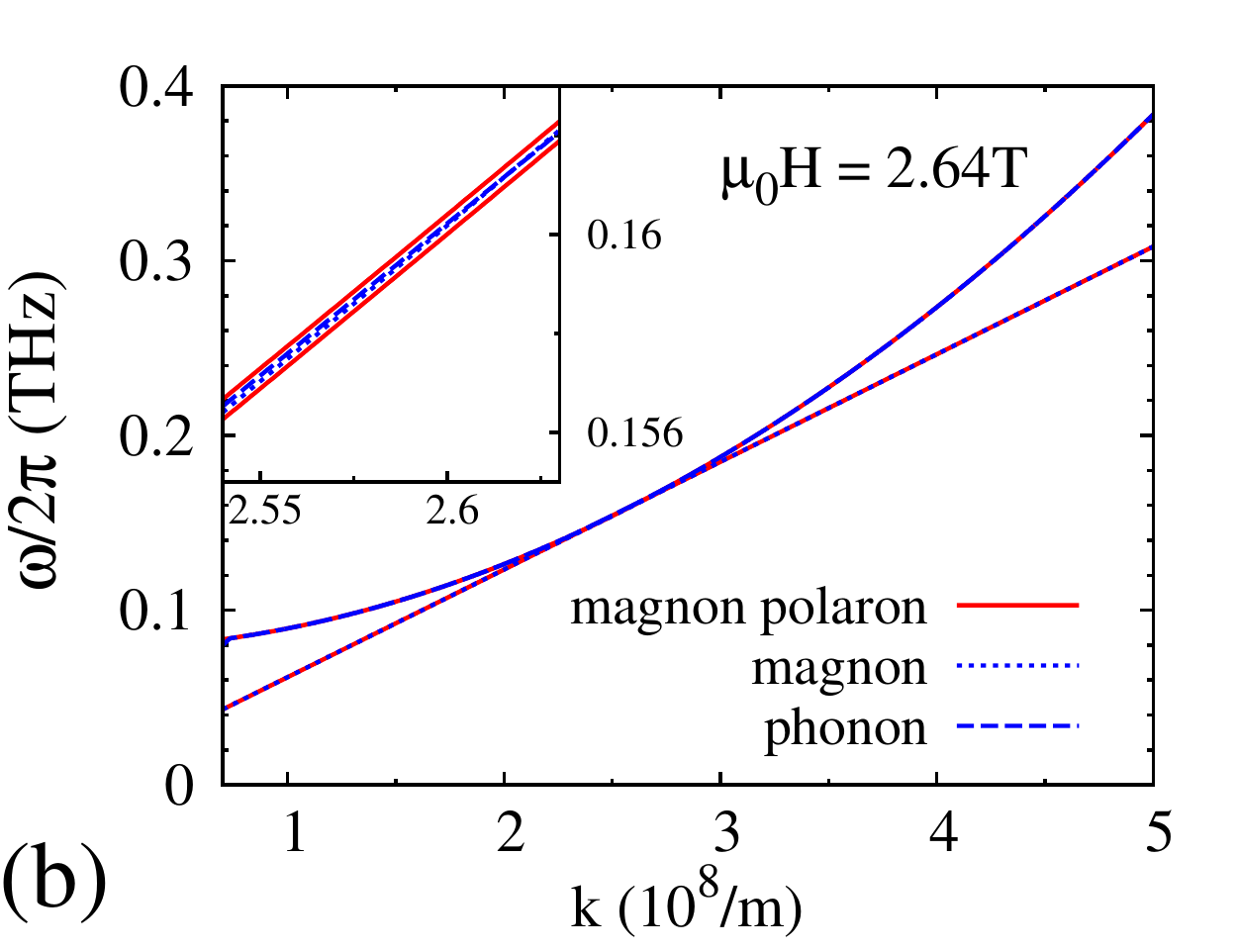}
\caption{Magnon, TA phonon ($\lambda=1$), and magnon-polaron mode dispersions for
YIG (see Table \ref{tab:table1} for parameters) with $\mathbf{H}\parallel\hat{\mathbf{z}}$ and
$\mathbf{k}\parallel\hat{\mathbf{x}}$ ($\theta=\pi/2$ and $\phi=0$). (a) For
$\mu_{0}H=1$\thinspace T, the magnon and transverse phonon dispersions
intersect at two crossing points $\text{k}_{1,2}$. The mixing between magnons
and phonons (see insets) is maximized at these crossings. (b) For $\mu
_{0}H_{\bot}=2.64$\thinspace T, the phonon dispersion becomes a tangent to the
magnon dispersion which maximizes the phase space of magnon-polaron formation
(see inset).}%
\label{1DDIS}%
\end{figure}

\section{Magnon-polarons}

Here we introduce magnon-polarons and formulate their semiclassical transport properties.

\subsection{Magnon-polaron modes}

We rewrite the Hamiltonian $\mathcal{H}=\mathcal{H}_{\mathrm{mag}}%
+\mathcal{H}_{\mathrm{el}}+\mathcal{H}_{\mathrm{mec}}$ as
\begin{equation}
\mathcal{H}=\frac{1}{2}\sum_{{\mathbf{k}}}%
\begin{bmatrix}
\boldsymbol{\beta}_{{\mathbf{k}}}^{\dagger} & \boldsymbol{\beta}%
_{-{\mathbf{k}}}%
\end{bmatrix}
\cdot\boldsymbol{H}_{{\mathbf{k}}}\cdot%
\begin{bmatrix}
\boldsymbol{\beta}_{{\mathbf{k}}} & \boldsymbol{\beta}%
_{-{\mathbf{k}}}^\dag%
\end{bmatrix}^T
\label{EQUAT18}%
\end{equation}
where $\boldsymbol{\beta}_{{\mathbf{k}}}^{\dagger}\equiv\left(  \alpha
_{{\mathbf{k}}}^{\dagger}\;c_{1{\mathbf{k}}}^{\dagger}\;c_{2{\mathbf{k}}%
}^{\dagger}\;c_{3{\mathbf{k}}}^{\dagger}\right)  $ and the Bogoliubov-de
Gennes Hamiltonian $\boldsymbol{H}_{{\mathbf{k}}}$ is an $8\times8$ Hermitian
matrix. 
Following Ref.~\cite{colpa78}, we introduce the para-unitary matrix
$\boldsymbol{\mathcal{T}}_{{\mathbf{k}}}$ that diagonalizes $\boldsymbol{H}%
_{{\mathbf{k}}}$ as
\begin{equation}
\boldsymbol{H}_{{\mathbf{k}}}\boldsymbol{\mathcal{T}}_{{\mathbf{k}}%
}=\boldsymbol{\nu}\boldsymbol{\mathcal{T}}_{{\mathbf{k}}}%
\begin{bmatrix}
\boldsymbol{E}_{{\mathbf{k}}} & \mathbf{0}\\
\mathbf{0} & -\boldsymbol{E}_{{-\mathbf{k}}}%
\end{bmatrix}
,
\end{equation}
where $[\boldsymbol{\nu}]_{jm}=\delta_{jm}\nu_{j}$ with $\nu_{j}=+1$ for
$j=1,..,4$ and $\nu_{j}=-1$ for $j=5,..,8$, and $\boldsymbol{E}_{{\mathbf{k}}%
}$ is a diagonal matrix, whose $i$-th element $\hbar\Omega_{i\mathbf{k}}$
represents the dispersion relation of the hybrid mode with creation operator
$\Gamma_{i{\mathbf{k}}}^{\dagger}=\sum_{j=1}^{8}[\boldsymbol{\beta
}_{{\mathbf{k}}}^{\dagger}\;\boldsymbol{\beta}_{-{\mathbf{k}}}]_{j}%
(\boldsymbol{\mathcal{T}}_{{\mathbf{k}}}^{-1})_{ij}^{\ast}$ that is neither
a pure phonon or  magnon, but a magnon-polaron.

Let us focus our attention to waves propagating perpendicularly to the
magnetic field, i.e., $\mathbf{k}=k\hat{\mathbf{x}}$ (see Fig.~\ref{scheme}). It follows from
Eq.~(\ref{hamiltME}) that magnon-polarons involve only TA phonons.
Disregarding the dipolar interactions, the TA phonon branch is tangent to the
magnon dispersion for $\mu_{0}H_{\bot}=c_{\bot}^{2}/4D_{ex}\gamma$ at
$k_{\bot}=c_{\bot}/2D_{ex}$. This estimate holds for $M_{s}\ll H_{\bot}$;
otherwise the dipolar interaction shifts the magnon dispersion to higher
values, leading to a smaller critical field $H_{\bot}$. For $H<H_{\bot}$, the
TA phonon dispersion intersects the spin wave spectrum at two crossing points,
$k_{1}$ and $k_{2}$,
\begin{equation}
k_{1,2}=k_{\bot}\mp\sqrt{k_{\bot}^{2}-\frac{\gamma\mu_{0}H}{D_{ex}}}\,,
\end{equation}
where the minus (plus) corresponds to the label 1 (2). In the vicinity of
$k_{1,2}$, the modes corresponding to the dispersions $\Omega_{1,2k}$ are strongly coupled, as shown in the inset of
Fig.~\ref{1DDIS}(a). The magnetoelastic coupling changes the crossing at
$k_{1,2}$ into an anti-crossing with energy splitting $\Delta\Omega_{k_{1,2}%
}=\Omega_{2k_{1,2}}-\Omega_{1k_{1,2}}$. For $k\ll k_{1}$, the $\Gamma_{1{k}%
}^{\dagger}$ ($\Gamma_{2{k}}^{\dagger}$) mode resembles closely  a pure lattice
vibration (spin wave) whilst for $k_{1}\ll k\ll k_{2}$ these roles are 
reversed, returning to their original character for $k\gg k_{2}$. At the
critical magnetic field $H_{\bot}$, the magnon dispersion shifts
upwards such
that the TA phonon branch becomes tangential. Figure ~\ref{1DDIS}(b) shows
that this \textquotedblleft touching\textquotedblright\ condition generates
the strongest effects of the MEC, since the magnon and phonon modes
are strongly coupled over a relatively large volume in momentum space. At
higher magnetic fields, the uncoupled magnonic and TA phononic curves no
longer cross, hence the MEC does not play a significant role, and
$\boldsymbol{\mathcal{T}}_{k}$ reduces to the identity matrix.

An analogous physical picture holds when considering the magnon-polaron modes 
arising from the coupling between magnons and LA phonons for $\sin 2\theta_{\mathbf k}\ne 0$, 
with critical field
$\mu_{0}H_{\parallel}=c_{\parallel}^{2}/4D_{ex}\gamma$ and touch point
$k_{\parallel}=c_{\parallel}/2D_{ex}$ (for $M_{s}\ll H_{\parallel}$).

\subsection{Magnon-polaron transport}

We proceed to assess the magnetoelastic coupling effects on the transport
properties of a magnetic insulator in order to model the spin Seebeck effect
and magnon injection by heavy metal contacts.

A non-equilibrium state at the interface between the magnetic insulator and
the normal metal generates a spin current that can be detected by the inverse
spin Hall effect, as shown in Fig.~\ref{scheme}. The spin current and spin-mediated heat currents are then
proportional to the interface spin mixing conductance that is governed by the
exchange interaction between conduction electrons in the metal and the
magnetic order in the ferromagnet. In the presence of magnon-polarons, the
excitations at the interface have mixed character. Since the spin-pumping and
spin torque processes are mediated by the exchange interaction, only the
magnetic component of the magnon-polaron in the metal interacts with the
conduction electrons. We focus here on the limit in which the smaller of the
magnon spin diffusion length and magnetic film thickness is sufficiently large
such that the spin current is dominated by the bulk transport and the
interface processes may be disregarded. We therefore calculate in the
following the spin-projected angular momentum and heat currents in the bulk of
the ferromagnet, assuming that the interface scattering processes and
subsequent conversion into an inverse spin Hall voltage do not change the
dependence of the observed signals on magnetic field, temperature gradient,
material parameters, etc..
\begin{table}[h]
\caption{\label{tab:table1}%
Selected YIG parameters~\cite{Gilleo58,Harris63,Manuilov09,Srivast73,Gurevich96,Strauss67,Eggers63,Hansen73}.
}
\begin{ruledtabular}
\begin{tabular}{ l c c c } 
\textrm{}&
\textrm{Symbol}&\textrm{Value}&\textrm{Unit} \\
\colrule 
Macrospin & S & 20 & -\\
g-factor & g& 2 & -\\
Lattice constant & $a_{0}$ & 12.376 & \text{\AA}\\
Gyromagnetic ratio & $\gamma$ & $2\pi\times28$ & GHz/T \\
Saturation magnetization & $\mu_{0} M_{s}$ & 0.2439 & \textrm{T} \\
Exchange stiffness & $D_{ex}$  &  $7.7 \times 10^{-6}$ & $\text{m} ^{2}/\text{s}$\\
LA-phonon sound velocity & $c_{\parallel} $ & $7.2\times10^{3}$&\text{m}/\text{s} \\
TA-phonon sound velocity & $c_{\bot} $ & $3.9\times10^{3}$&\text{m}/\text{s} \\
Magnetoelastic coupling  & $B_{\bot}$ & $2\pi \times 1988$ & GHz \\
Average mass density & $\bar{\rho} $ & $5.17 \times 10^{3}$ &  $\text{Kg}/ \text{m}^{3}$ \\
Gilbert damping & $\alpha$ & $10^{-4}$ & -
\end{tabular}
\end{ruledtabular}
\end{table}
Since the phonon specific heat is an order of magnitude larger than the magnon
one at low temperatures~\cite{Boona14}, we may assume that the phonon
temperature and distribution is not significantly perturbed by the magnons.
$T$ is the phonon temperature at equilibrium and we are interested in the
response to a constant gradient $\boldsymbol{\nabla}T\Vert\mathbf{\hat{x}}$.
The spin-conserving relaxation of the magnon distribution towards the phonon
temperature is assumed to be so efficient that the magnon temperature is
everywhere equal to the phonon temperature. Also the magnon-polaron
temperature profile is then $T(x)=T+\left\vert \boldsymbol{\nabla}T\right\vert x$.
Assuming efficient thermalization of both magnons and phonons and weak
spin-non-conserving processes as motivated by the small Gilbert damping, a
non-equilibrium distribution as injected by a metallic contact can be
parameterized by a single parameter, viz. the effective magnon-polaron
chemical potential $\mu$~\cite{ludo}. This approximation might break down at a
very low temperatures, but to date there is no evidence for that.

In equilibrium the chemical potential of magnons and phonons vanishes since
their number is not conserved. The occupation of the $i$-th magnon-polaron
in equilibrium is therefore given by the Planck distribution function
\begin{equation}
f_{i{\mathbf{k}}}^{(0)}=\left(  \exp\frac{\hbar\Omega_{i\mathbf{k}}}{k_{B}%
T}-1\right)  ^{-1}\,.
\end{equation}
Note that here we have assumed the $i$-th magnon polaron scattering rate to be
sufficiently smaller than the gap between the magnon-polaron mode
dispersions, i.e.,  $\tau_{i\mathbf{k}_{i}}^{-1}\ll\Delta\Omega_{\mathbf{k}%
_{i}}$ for every $\mathbf{k}_{i}$, which guarantees the $i$-th magnon-polaron to not dephase and hence its distribution function  to be well-defined.
We focus on films with thickness $L\gg\Lambda_{\mathrm{mag}},\Lambda
_{\mathrm{ph},\lambda}, \ell_{\text{m}}, \ell_{\text{ph}, \lambda}$, where $\Lambda_{\mathrm{mag}}=(4\pi\hbar D_{ex}%
/k_{B}T)^{1/2}$ and $\Lambda_{\mathrm{ph},\lambda}=\hbar c_{\lambda}/k_{B}T$ are the
thermal magnon and phonon (de Broglie) wavelengths, respectively, and $\ell_{\text{m}}$ ($\ell_{\text{ph}, \lambda}$) the magnon (phonon) mean free path. The bulk
transport of magnon-polarons is then semiclassical and can be treated by means
of Boltzmann transport theory. In the relaxation time approximation to the
collision integral, the Boltzmann equation for the out-of-equilibrium
distribution function $f_{i{\mathbf{k}}}(\mathbf{r},t)$ reads
\begin{equation}
{\partial_{t}f_{i{\mathbf{k}}}}+{\partial_{\mathbf{r}}}{f_{i{\mathbf{k}}}%
}\cdot{\partial_{\mathbf{k}}\Omega_{i{\mathbf{k}}}}=-({f_{i{\mathbf{k}}%
}-f_{i{\mathbf{k}}}^{(0)}})/{\tau_{i\mathbf{k}}}\,, \label{BOLTZ}%
\end{equation}
where $\tau_{i\mathbf{k}}$ is the relaxation time towards equilibrium. In the
steady state, the deviation $\delta f_{i{\mathbf{k}}}(\mathbf{r}%
)=f_{i{\mathbf{k}}}(\mathbf{r})-f_{i{\mathbf{k}}}^{(0)}$ encodes the magnonic
spin, $\mathbf{j}_{\mathrm{m}}$, and heat, $\mathbf{j}_{Q,\mathrm{m}}$,
current densities
\begin{align}
\mathbf{j}_{\mathrm{m}}  &  =\int\frac{d^{3}\mathbf{k}}{(2\pi)^{3}}\sum
_{i}W_{i\mathbf{k}}(\partial_{\mathbf{k}}\Omega_{i\mathbf{k}})\delta
f_{i\mathbf{k}}\,,\label{EQUAZ18}\\
\mathbf{j}_{Q,\mathrm{m}}  &  =\int\frac{d^{3}\mathbf{k}}{(2\pi)^{3}}\sum
_{i}W_{i\mathbf{k}}(\partial_{\mathbf{k}}\Omega_{i\mathbf{k}})(\hbar
\Omega_{i\mathbf{k}})\delta f_{i\mathbf{k}}\,. \label{EQUAZ19}%
\end{align}
Here, $W_{i\mathbf{k}}=|(\boldsymbol{U}_{\mathbf{k}})_{i1}|^{2}%
+|(\boldsymbol{U}_{\mathbf{k}})_{i5}|^{2}$ is the magnetic amplitude of the
$i$-th quasi-particle branch with $\boldsymbol{U}_{\mathbf{k}}%
=\boldsymbol{\mathcal{T}}_{\mathbf{k}}^{-1}$. For small temperature 
gradients, Eqs.~(\ref{EQUAZ18}) and~(\ref{EQUAZ19}) can be linearized
\begin{align}
\mathbf{j}_{\mathrm{m}}  &  \simeq -\boldsymbol{\sigma}\cdot\boldsymbol{\nabla}%
\mu-\boldsymbol{\zeta}\cdot\boldsymbol{\nabla}{T}\,,\label{EQZ21}\\
\mathbf{j}_{Q,\mathrm{m}}  &  \simeq -\boldsymbol{\rho}^{(\mathrm{m})}\cdot
\boldsymbol{\nabla}\mu-\boldsymbol{\kappa}^{(\mathrm{m})}\cdot\boldsymbol{\nabla
}{T}\,, \label{EQ22}%
\end{align}
where the tensors $\boldsymbol{\sigma}$, $\boldsymbol{\kappa}^{(\mathrm{m})}$ ,
$\boldsymbol{\zeta}$, and $\boldsymbol{\rho}^{(\mathrm{m})}(=T\boldsymbol{\zeta
}$ by the Onsager-Kelvin relation)  are, respectively, the spin and (magnetic) heat
conductivities, and the spin Seebeck and Peltier coefficients. In the absence
of magnetoelastic coupling, Eqs.~(\ref{EQZ21}) and~(\ref{EQ22}) reduce to the
spin and heat currents of magnon diffusion theory~\cite{ludo}.

The total heat current $\mathbf{j}_{Q}$ carried by both magnon and phonon
systems does not invoke the spin projection $W_{i\mathbf{k}}$, i.e.,
\begin{align}
\mathbf{j}_{Q}  &  =\int\frac{d^{3}\mathbf{k}}{(2\pi)^{3}}\sum_{i}%
(\partial_{\mathbf{k}}\Omega_{i\mathbf{k}})(\hbar\Omega_{i\mathbf{k}})\delta
f_{i\mathbf{k}}\,,\nonumber\\
&  \simeq-\boldsymbol{\kappa}\cdot\boldsymbol{\nabla}{T}\,,
\end{align}
where $\boldsymbol{\kappa}$ is the total heat conductivity.

In terms of the general transport coefficients
\begin{align}
L_{\alpha\gamma}^{mn}  &  =\beta\int\frac{d^{3}\mathbf{k}}{(2\pi)^{3}}\sum
_{i}(W_{i\mathbf{k}})^{m}\tau_{i\mathbf{k}}(\partial_{k_{\alpha}}%
\Omega_{i\mathbf{k}})(\partial_{k_{\gamma}}\Omega_{i\mathbf{k}})\nonumber\\
&  \times\frac{e^{\beta\hbar\Omega_{i\mathbf{k}}}}{(e^{\beta\hbar
\Omega_{i\mathbf{k}}}-1)^{2}}(\hbar\Omega_{i\mathbf{k}})^{n}\,,
\label{generaltensor}%
\end{align}
(with $\beta=1/k_{B}T$), we identify $\sigma
_{\alpha\gamma}=L_{\alpha\gamma}^{10}$, $\zeta_{\alpha\gamma}=L_{\alpha\gamma}^{11}/T$,
$\kappa_{\alpha\gamma}^{(\mathrm{m})}=L_{\alpha\gamma}^{12}/T$ and
$\kappa_{\alpha\gamma}=L_{\alpha\gamma}^{02}/T$.

At low temperatures, the excitations relax dominantly by elastic magnon- and
phonon-disorder scattering as modelled here by Eqs.~(\ref{Himpurity}) and
(\ref{Himpurityphonon}), respectively. The Fermi Golden Rule scattering rate
$\tau_{i\mathbf{k}}^{-1}$ of the $i$-th magnon-polaron reads
\begin{align}
\tau_{i\mathbf{k}}^{-1}  &  =\frac{2\pi}{\hbar}\sum_{l=1}^{4}\sum
_{j\mathbf{k}^{\prime}}\left[  (\boldsymbol{U}_{\mathbf{k}^{\prime}}%
)_{jl}^{\ast}(\boldsymbol{U}_{\mathbf{k}})_{il}\right. \nonumber\\
&  \left.  +(\boldsymbol{U}_{\mathbf{k}^{\prime}})_{jl+4}^{\ast}%
(\boldsymbol{U}_{\mathbf{k}})_{il+4}\right]  ^{2}|v_{l}|^{2}\delta(\hbar
\Omega_{i\mathbf{k}}-\hbar\Omega_{j\mathbf{k}^{\prime}})\,, \label{tauik}%
\end{align}
where $v_{1}=v^{\mathrm{mag}}$ and $v_{2,3,4}=v^{\mathrm{ph}}$, while the purely magnonic and phononic scattering rates are given by 
\begin{align}
\tau^{-1}_{\mathbf{k}, \text{mag}} =  \frac{L^{3}|v^{\text{mag}}|^{2}}{2\pi \hbar^{2} D_{ex}} k\,, \; \; \; \; \; \; \; \; \; \;  \tau^{-1}_{\mathbf{k}, \text{ph} _{\lambda}}=\frac{L^{3} |v^{\text{ph}}|^{2}}{\pi \hbar^{2} c_{\lambda} } k^{2} \,.
\end{align}

\section{Results}
\begin{figure}[ptb]
\includegraphics[width=6cm]{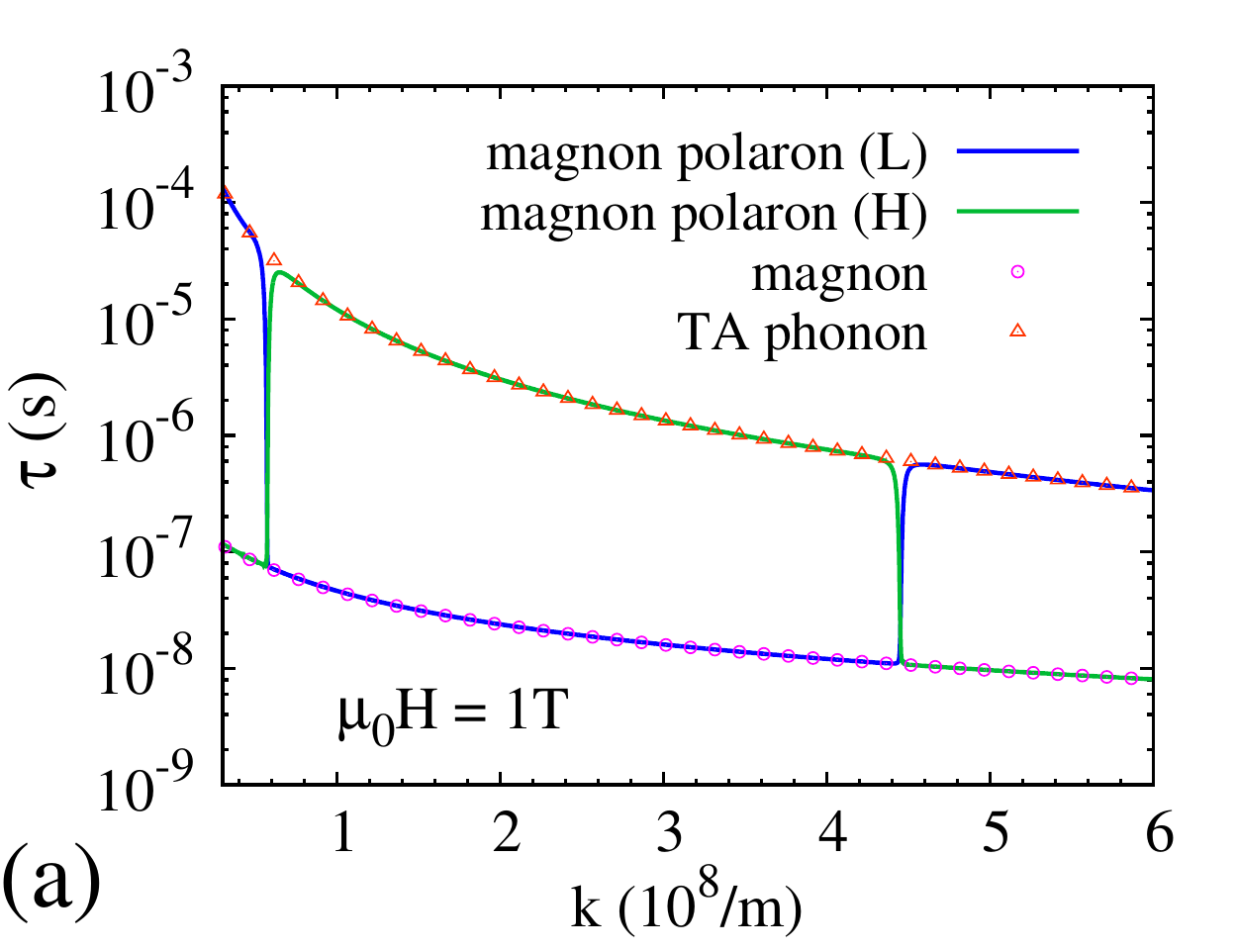}
\includegraphics[width=6cm]{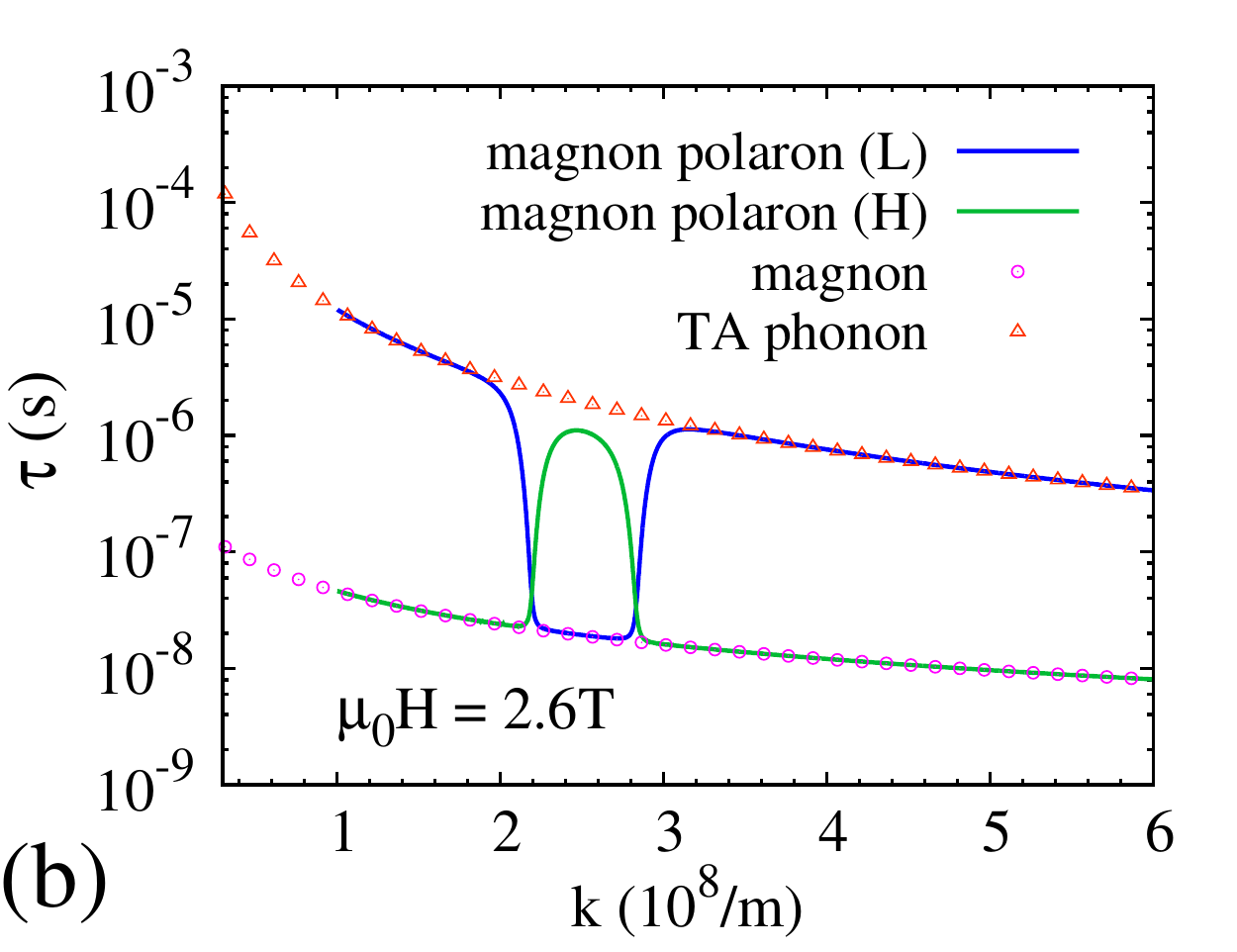}\caption{(a) Scattering times of
magnons, TA phonons ($\lambda=1$), 
and lower (L)/upper (H) branch magnon-polarons in YIG for $\mu_{0}%
H=1$\thinspace T ($\mathbf{H}\parallel\hat{\mathbf{z}}$) as a function of
wave vector $\mathbf{k}\parallel\hat{\mathbf{x}}$  for $\eta=100$. (b) Same as (a) but $\mu_{0}H_{\bot}=2.64$\thinspace
T. }%
\label{relaxationtime}%
\end{figure}

In this section we discuss our numerical results for the transport
coefficients, in particular the emergence of field and temperature
dependent anomalies, and we compare the thermally induced spin current with
measured spin Seebeck voltages~\cite{kikkawa}.

\subsection{Spin and heat transport}

We consider a sufficiently thick ($>1\,\mathrm{\mu}$m) YIG film subject to a
temperature gradient $\boldsymbol{\nabla}T\parallel\boldsymbol{\hat{x}}$ and
magnetic field $\mathbf{H}\parallel\boldsymbol{\hat{z}}$, as illustrated in Fig.~\ref{scheme}.  The parameters we employ are summarized in Table \ref{tab:table1}.
 A scattering potential
$|v^{\mathrm{mag}}|^{2}=10^{-5}$~$\mathrm{s}^{-2}$ (with $v^{\mathrm{mag}}$ in
units of $\hbar$) reproduces the observed low-temperature magnon mean free
path~\cite{Boona14}. We treat the ratio between magnetic and non-magnetic
impurity-scattering potentials, $\eta=|v^{\mathrm{mag}}/v^{\mathrm{ph}}|^{2}$,
as an adjustable parameter. With the deployed scattering potentials $\tau
_{\mathbf{k}_{i}}^{-1}\ll\Delta\Omega_{\mathbf{k}_{i}}$ for all magnon-polaron
modes, ensuring the validity of our treatment. We compute the integrals
appearing in Eq.~(\ref{generaltensor}) numerically on a fine grid ($ \sim 10^{6}$
$k$-points) to guarantee accurate results. 

Figure~\ref{relaxationtime}(a) shows the magnon-polaron scattering times and
how they deviate from the purely phononic and magnonic ones 
close to the anticrossings. At the \textquotedblleft
touching\textquotedblright\ fields the phase space portion over which the scattering
times are modified with respect to the uncoupled situation is maximal (see Fig.2(b)) as are the
effects on spin and heat transport properties  as
discussed below. 

\begin{figure}[ptb]
\includegraphics[width=6cm]{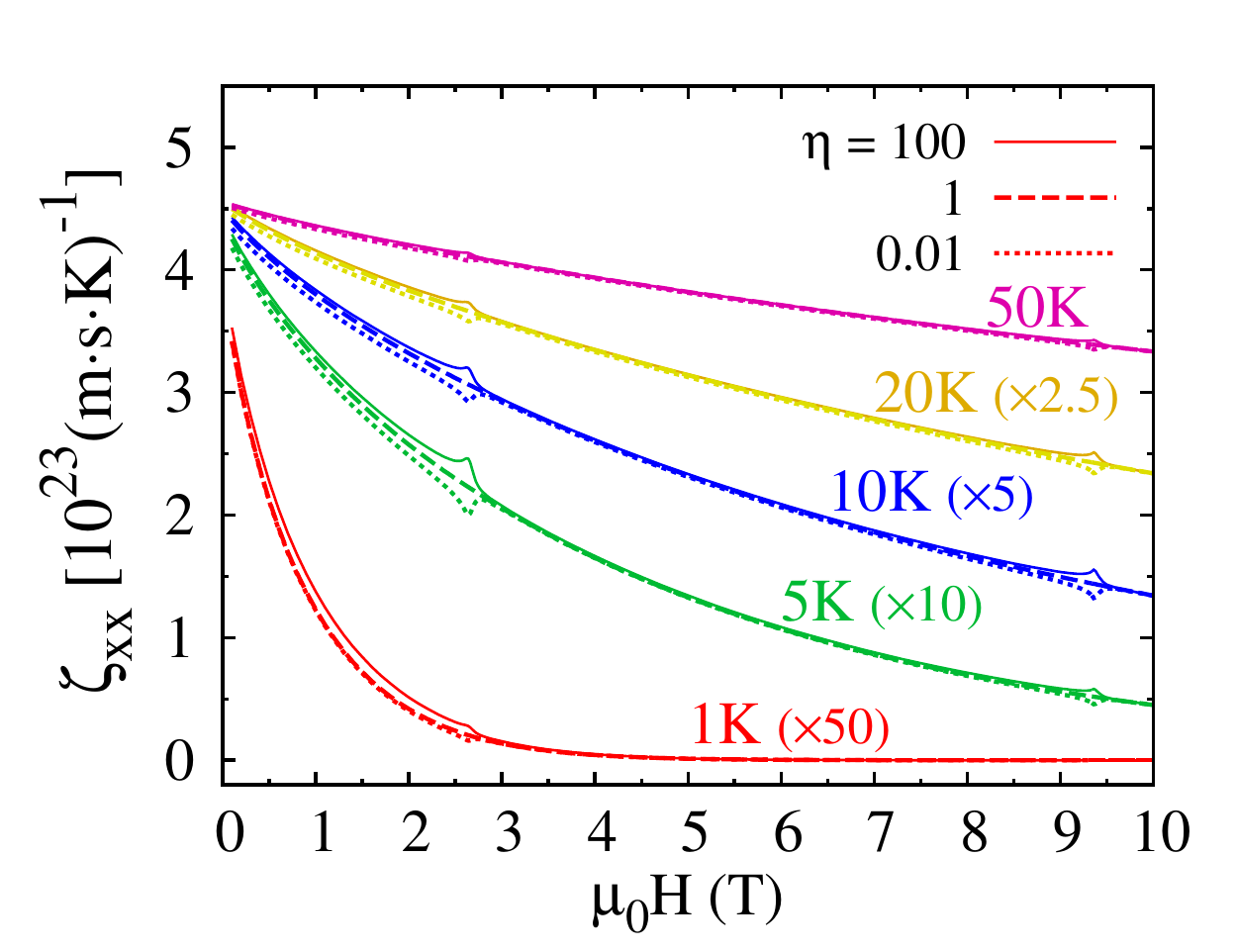} \caption{Magnetic field and temperature
dependence of the spin Seebeck coefficient $\zeta_{xx}$ for
different values of the ratio $\eta$ between magnon and phonon
impurity-scattering potentials.}%
\label{3DBT}%
\end{figure}

\begin{figure}[ptb]
\includegraphics[width=6cm]{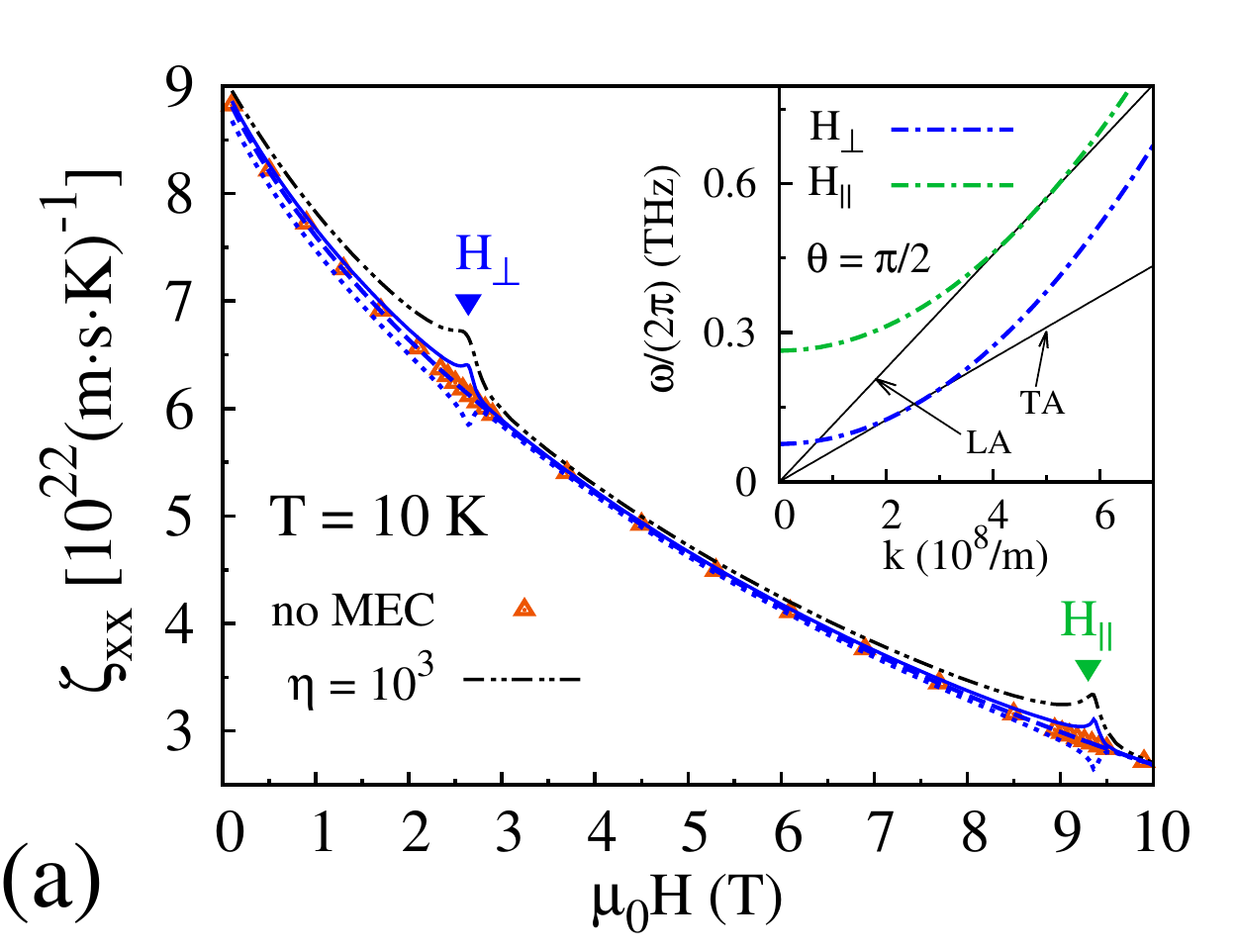} 
\includegraphics[width=6cm]{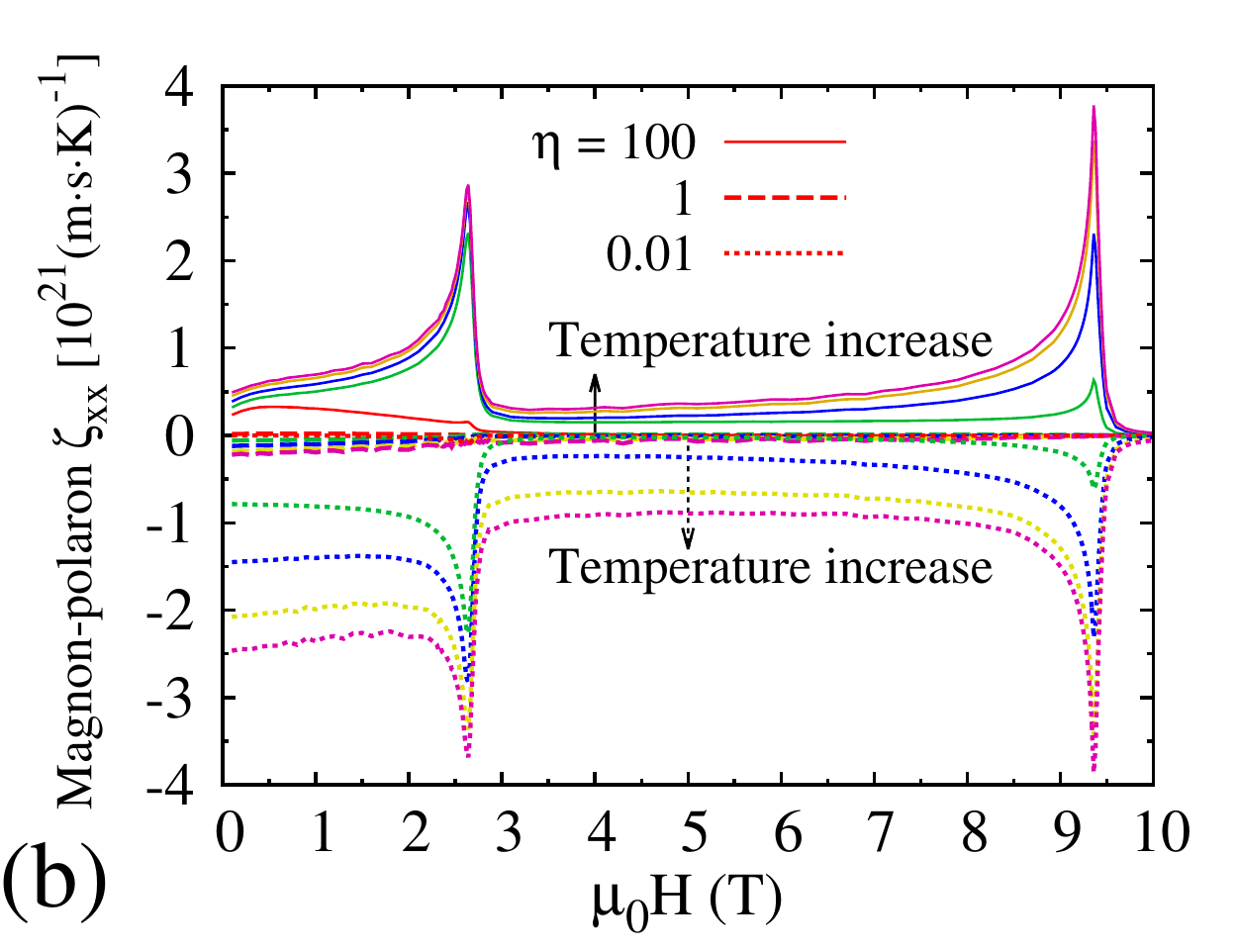}
\includegraphics[width=6cm]{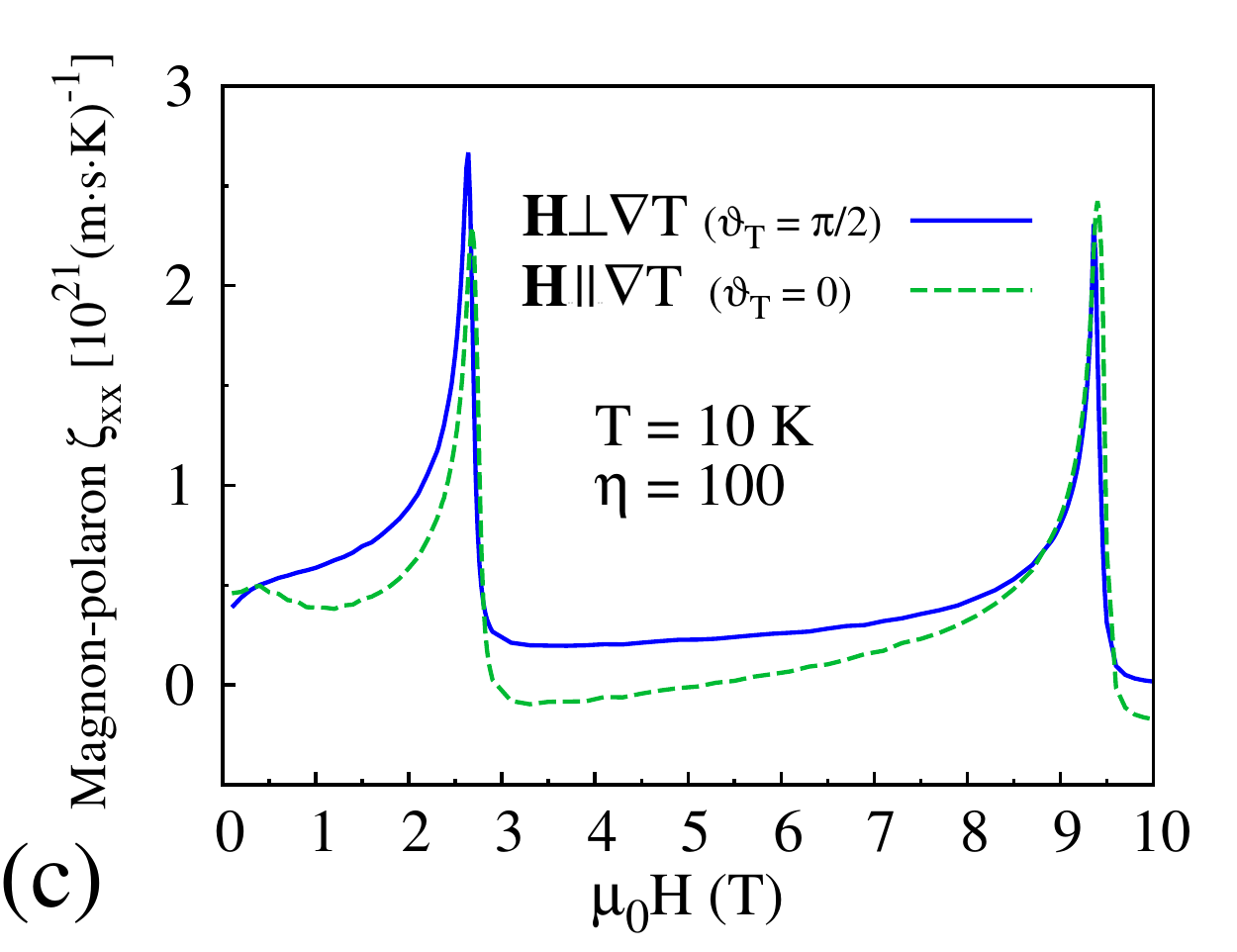} 
\caption{(a) Spin Seebeck coefficient
$\zeta_{xx}$ of bulk YIG as a function of magnetic field at $T=10$ K. The
black dash-dotted, blue solid, blue dashed, blue dotted lines are computed
for, respectively, $\eta=10^{3},100,1,0.01$. The triangles are obtained for
zero MEC. The inset shows the dispersions of uncoupled transverse (TA) and
longitudinal (LA) acoustic phonons and the magnons shifted by $H_{\parallel}$
and $H_{\bot}$ magnetic fields. (b) The magnetic field and temperature
dependence of the magnon-polaron contribution for different values of the
ratio $\eta$ between magnon and phonon impurity-scattering potentials.
(c)
$\zeta_{xx}$ as function of magnetic field for $\mathbf{H}\perp
\boldsymbol{\nabla}T$ (blue solid line) and $\mathbf{H}\Vert\boldsymbol{\nabla
}T$ (green dashed line) at $T=10$ K for $\eta=100$.}%
\label{3D5K}%
\end{figure}

In Fig.~\ref{3DBT}, we plot the (bulk) spin Seebeck coefficient $\zeta
_{xx}$ as a function of magnetic field for different values of $\eta$. For
$\eta=1$, $\zeta_{xx}$ decreases monotonously with increasing magnetic field,
while for $\eta\neq1$ two anomalies are observed at $\mu_{0}H_{\bot}%
\sim2.64\,$T and $\mu_{0}H_{\parallel}\sim9.3\,$T. More precisely, peaks
(dips) appear for $\eta=100(0.01)$ at the same magnetic fields but with
amplitudes that depend on temperature.
The underlying physics can be understood in terms of the
dispersion curves plotted in the inset of Fig.$\,$\ref{3D5K}(a). The first
(second) anomaly occurs when the TA (LA) phonon branch becomes a tangent of
the magnon dispersion, which maximizes the integrated magnon-polaron coupling.
The group velocity of the resulting magnon-polaron does not differ substantially
from the purely magnonic one, but its scattering time can be drastically
modified, depending on the ratio between the magnonic and phononic scattering potentials
(see Fig.~\ref{relaxationtime}(b)). The spin currents can
therefore be both enhanced or suppressed by the MEC. When the magnon-impurity
scattering potential is larger than the phonon-impurity one, the hybridization
induced by the MEC lowers the effective potential perceived by magnons, giving
rise to an enhanced scattering time and hence larger currents. This can be
confirmed by comparing the blue 
solid ($\eta=100$) and the black dash-dotted ($\eta=10^{3}$)
lines in Fig.~\ref{3D5K}(a), showing that the magnitude of the peaks
increases with increasing $\eta$. 
When magnetic and non-magnetic scattering
potentials are the same, i.e., $\eta=1$, the anomalies vanish as illustrated
by the dashed blue line in Fig.$\,$\ref{3D5K}(a), and agrees with the results obtained in the
absence of MEC (triangles).

The frequencies at which magnon and phonon dispersions are tangential for
uncoupled transverse and longitudinal modes are 0.16~THz $\left(  \hat
{=}8~\mathrm{K}\right)  $and 0.53~THz $\left(  \hat{=}26~\mathrm{K}\right)  $.
Far below these temperatures, the magnon-polaron states are not populated, which
explains the disappearance of the second anomaly and the strongly reduced
magnitude of the first one at 1\thinspace K in Fig.$\,$\ref{3DBT}.
In the opposite limit, the higher energy anomaly becomes relatively stronger [see the
solid curve at 50~K in Fig.$\,$\ref{3DBT}].
The overall decay of the spin Seebeck coefficient with increasing magnetic
field is explained by the freeze-out caused by the increasing magnon gap
opened by the magnetic field [see the inset of Fig.$\,$\ref{3D5K}(a)].
\begin{figure}[ptb]
\includegraphics[width=6cm]{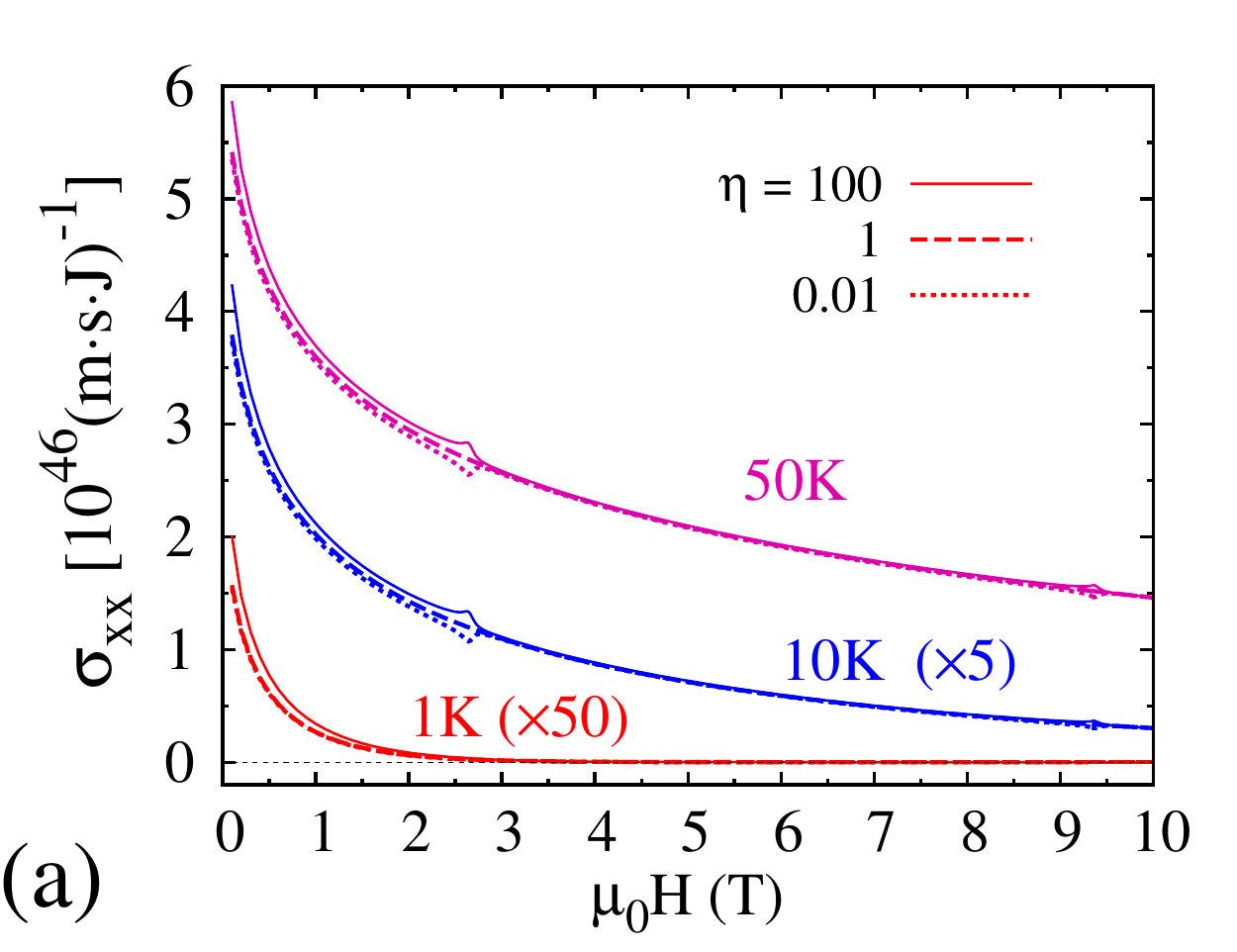}
\includegraphics[width=6cm]{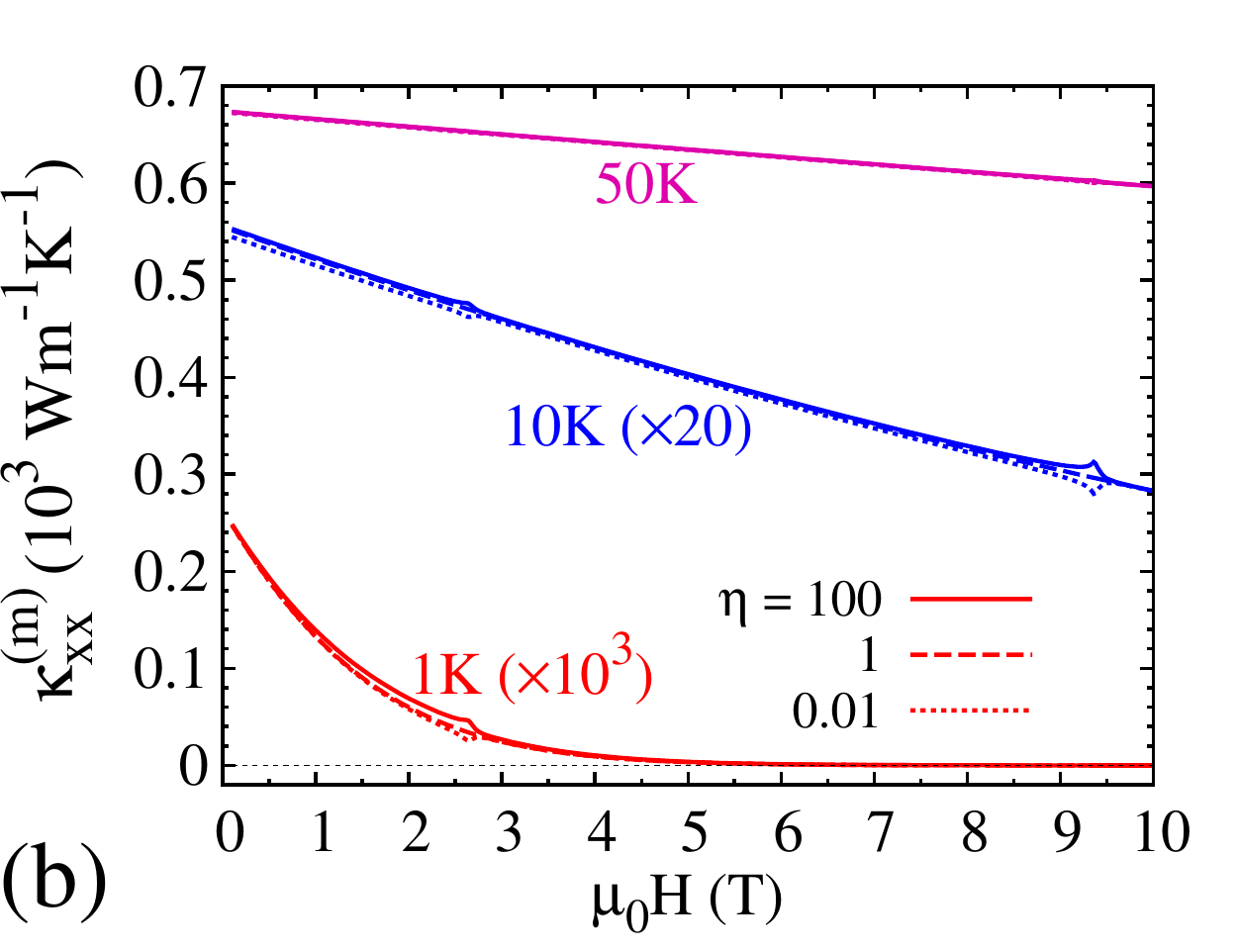}\caption{(a) The magnetic field and
temperature dependence of the magnon spin conductivity $\sigma_{xx}$ for
different values of the ratio $\eta$ between magnon and phonon
impurity-scattering potentials. (b) The magnetic field and temperature
dependence of the magnon heat conductivity $\kappa_{xx}^{(\text{m})}$ for
different values of the ratio $\eta$ between magnon and phonon
impurity-scattering potentials.}%
\label{othercoefficients}%
\end{figure}This strong decrease has been observed in single YIG
crystals~\cite{Kikkawa15, JinPRB15}, but it is suppressed in thinner samples or even
enhanced at low temperatures~\cite{kikkawa}. The effect is tentatively
ascribed to the paramagnetic GGG substrate that becomes magnetically active a
low temperatures~\cite{kikkawa} and is beyond the scope of the present theory.
We therefore subtract the pure magnonic background (triangles in
Fig.~\ref{3D5K}(a)) from the magnon-polaron spin currents, which leads to the
net magnon-polaron contribution 
shown in Fig.$\,$\ref{3D5K}(b).

The dipolar interaction is responsible for the anisotropy in the magnon
dispersion in Eq.~(\ref{magnond}), which is reflected in the magnetic field dependence
of the heat and spin currents. In Fig.$\,$\ref{3D5K}(c) we plot $\zeta_{xx}$
as function of the angle $\vartheta_{T}$ between magnetic field and transport
direction for $\eta=100$ and $T=10$ K. The magnon-polaron contributions for
magnetization parallel and perpendicular to the transport are plotted as the
green dashed and blue solid curves, respectively. The  anisotropy
shifts the magnon-polaron peak positions, but does not substantially modify
their amplitude. On these grounds, we proceed with computing other transport
coefficients for the configuration $\mathbf{H}\bot\boldsymbol{\nabla}T$ only.

Figure~\ref{othercoefficients}(a) shows the magnon spin conductivity
$\sigma_{xx}$ as function of the magnetic field and temperature for different
values of $\eta$. Two peaks (dips) appear at $H_{\bot}$ and $H_{\parallel}$
for $\eta=100$ ($\eta=0.01$) at $10$ K and $50$ K, while they disappear for
$\eta=1$. At very low temperatures, $T=1$\thinspace K, the anomalies are not
visible anymore. The dependence of the spin conductivity on the temperature,
on the angle between the magnetic field and temperature gradient, and on the
scattering potentials ratio $\eta$ is the same as reported for the spin Seebeck coefficient $\zeta_{xx}$.

In Fig.~\ref{othercoefficients}(b), we plot the dependence of the magnon heat
conductivity $\kappa_{xx}^{(\mathrm{m})}$ on the magnetic field and on the temperature for
different values of $\eta$. The only difference with respect to the coefficient $\zeta_{xx}$ is
in the ratio between the amplitudes of the two anomalies at $T=10$ K, at which
the magnon modes contributing to the low-field ($H_{\bot})$ anomaly are
thermally excited, in contrast to high field $\left(  H_{\parallel}\right)  $
modes. In $\zeta_{xx}$ the anomaly at $H_{\bot}$ should therefore by better
visible, as is indeed the case. The magnon heat conductivity from
Eq.~(\ref{generaltensor}) contains an additional factor in the integrand which
is proportional to the energy of the magnon-polaron modes. The latter
compensates for the lower thermal occupation, which explains why the  anomaly at $H_{\parallel}$ is more pronounced in comparison with the spin Seebeck effect.

Perhaps surprisingly, the total heat conductivity $\kappa_{xx}$ in Fig.~\ref{fig6}(a) displays only
\emph{dips} for $\eta\neq1$ at the special fields $H_{\bot,\parallel}$. This can be explained as follows.  For
$\eta \gg 1$, the phonon contribution to the heat conductivity is larger than
the magnon contribution. Except at the critical
fields $H_{\bot,\parallel}$, the magnetic field dependence of $\kappa_{xx}$ is
therefore very weak (solid blue line). When phonons mix with magnons with a
short scattering time, the thermal conductivity is suppressed, causing the
dips close to $H_{\bot,\parallel}$. For $\eta \ll 1$, on the other hand, the
magnon contribution to heat conductivity prevails, as is seen by the strong
magnetic field dependence of $\kappa_{xx}$ (dotted blue line). Since now
$|v^{\mathrm{mag}}|<|v^{\mathrm{ph}}|$, the heat conductivity of the resulting
magnon-polaron mode is lower than the purely magnonic one. Again dips appear
close to the \textquotedblleft touching\textquotedblright\ magnetic fields.

Experimentally, the magnon heat conductivity $\kappa_{xx}^{(\mathrm{m,exp})}$ 
at a given temperature was referred to the difference between finite-field 
value $\kappa_{xx} (H)$ and $\kappa_{xx}(\infty)$, 
{i.e., $\kappa_{xx}^{(\mathrm{m,exp})}(H)=\kappa_{xx}(H)-\kappa_{xx}(\infty)$~\cite{Boona14}. The latter, $\kappa_{xx}(\infty)$, corresponds to the saturation value of the heat conductivity at high-field limit, above which it becomes a constant function of the magnetic field, suggesting that the magnon contribution has been completely frozen out and only the phonon
contribution remains. 
In general,  $\kappa_{xx}%
^{(\mathrm{m})}$ and $\kappa_{xx}^{(\mathrm{m,exp})}$ differ in the presence of
magnetoelasticity. The magnon heat conductivity $\kappa_{xx}^{(\mathrm{m,exp})}$
in Fig.~\ref{fig6}(b), evaluated by subtracting the high-field limit for $T=10$ K, shows dips for both $\eta=0.01$ and $\eta=100,$ in
contrast to the magnon heat conductivity $\kappa_{xx}^{(\mathrm{m})}$ in Fig.
\ref{othercoefficients}(b) with peaks for $\eta=100$.  The disagreement stems from $\kappa
_{xx}(\infty),$ which is the (pure) phonon contribution to the heat
conductivity at infinite magnetic fields, but is not the same as the phonon
heat conductivity at ambient magnetic fields when the MEC is significant. In
the latter case, the phonon heat conductivity itself depends on the magnetic
field and displays anomalies at $H_{\bot,\parallel}$; hence $\kappa
_{xx}^{(\mathrm{m,exp})}\neq\kappa_{xx}^{(\mathrm{m})}$.

\begin{figure}[ptb]
\includegraphics[width=6cm]{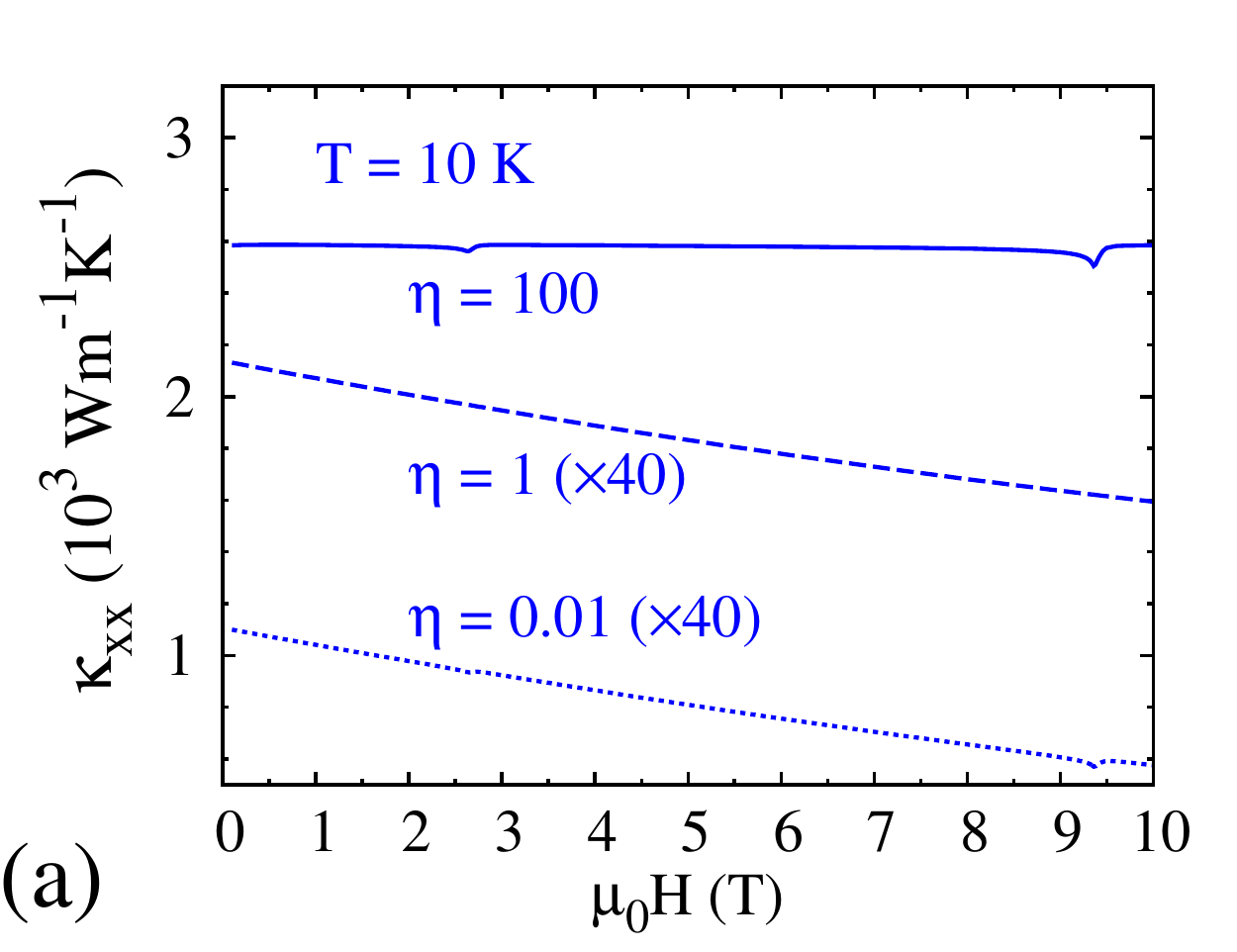}
\includegraphics[width=6cm]{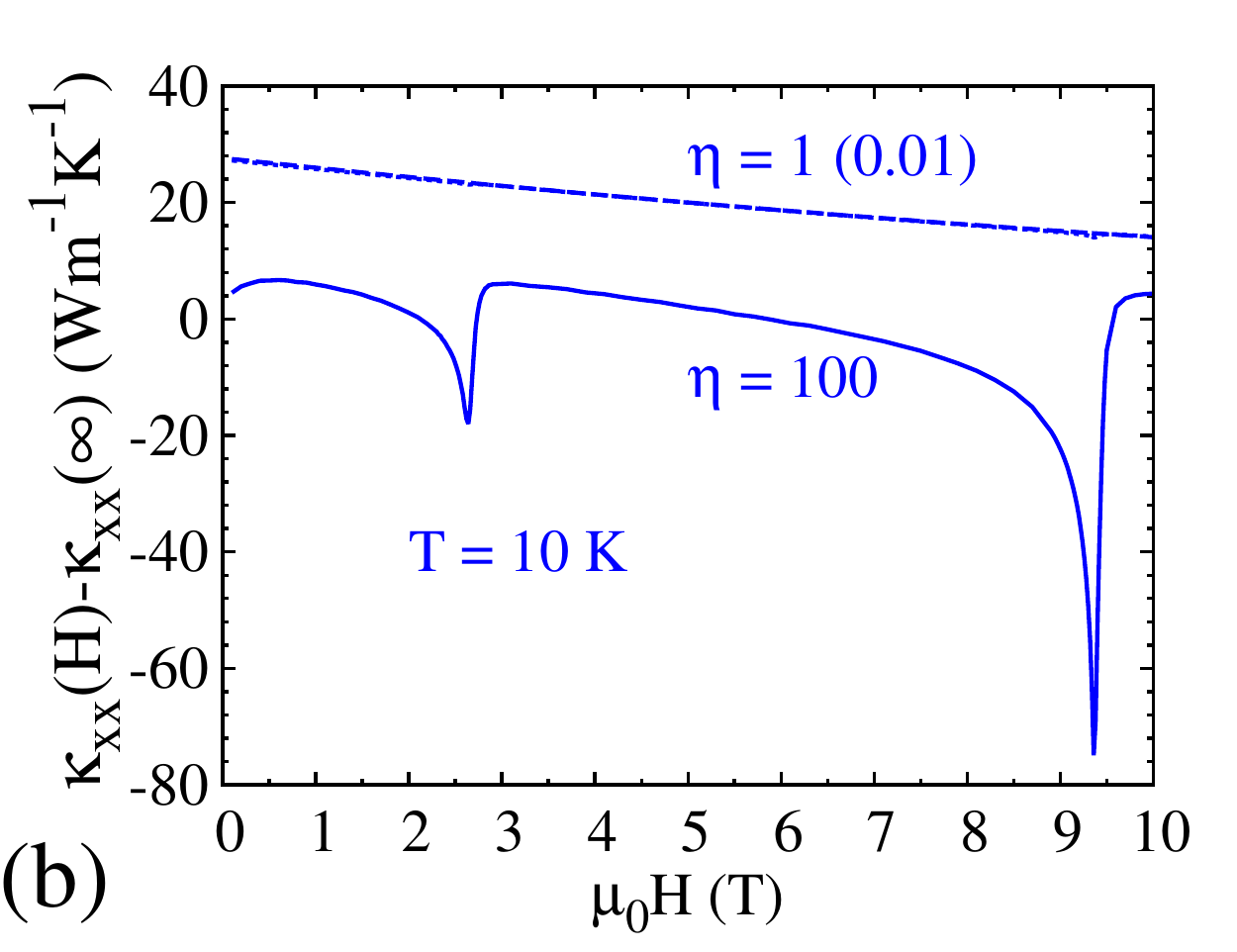}\caption{(a) The magnetic field
dependence of the heat conductivity $\kappa_{xx}$ at $T=10$ K for different
scattering parameters $\eta$. (b) Magnetic field dependence of the heat
conductivity difference $\kappa_{xx}\left(  H\right)  -\kappa_{xx}(\infty)$
simulating the experimental procedure~\cite{Boona14} at $T=10$ K.
}%
\label{fig6}%
\end{figure}

Nonetheless $\kappa_{xx}^{(\mathrm{m,exp})}$ can be useful since its fine
structure contains information about the ratio between the magnon-impurity and
phonon-impurity scattering potentials $|v^{\mathrm{mag}}|$ and
$|v^{\mathrm{ph}}|$. Also, $\kappa_{xx}\left(  \infty\right)  $ for $\eta=100$
is much larger than for $\eta=0.01$, and its value gives additional information
about the relative acoustic and magnetic quality of the sample.  
For example, the results reported by Ref.~\cite{Boona14} can be interpreted, within our theory, as suggesting a much higher acoustic than magnetic quality  of the samples, i.e., $\eta \gg 1$. 
The authors, however, have not investigated the magnetic field dependence of the heat conductivity but rather the temperature dependence, which is beyond the scope of this work. It is worth to mention that already the work of Ref.~\cite{Douglass1963} suggests that impurity scattering plays a key role in determining the  magnetic field dependence of the heat conductivity.

The appearance of the anomalies can be understood analytically with  few straightforward
simplifications. Let us consider a one-dimensional system along $\hat
{\mathbf{x}}$ and $\mathbf{H}=\left(  0,0,H\right)  .$ According to
Eq.~(\ref{hamiltME}) only the TA phonons couple to the magnons leading to the
magnon-polaron dispersion
\begin{equation}
\Omega_{1,2k}=\frac{\omega_{k}+\omega_{1k}\pm\sqrt{(\omega_{k}-\omega
_{1k})^{2}+\tilde{\omega}_{k}^{2}}}{2}\,,\label{polarondisp}%
\end{equation}
where $\tilde{\omega}_{k}=(S_{\bot}k)^{1/2}$ and $S_{\bot}=(nB_{\bot}%
)^{2}(\gamma \hbar^{2}/4M_{s}\bar{\rho}c_{\bot})$. The magnon-polaron spin amplitudes
$W_{1,2k}$ are
\begin{equation}
W_{1k}=\frac{\omega_{k}-\omega_{1k}+\sqrt{(\omega_{k}-\omega_{1k})^{2}%
+\tilde{\omega}_{k}^{2}}}{2\sqrt{(\omega_{k}-\omega_{1k})^{2}+\tilde{\omega
}_{k}^{2}}}\,,\label{31}%
\end{equation}
and $W_{2k}=1-W_{1k}$. Disregarding the small dipolar interactions ($M_{s}\ll
H_{\bot})$ the uncoupled dispersions touch at $\mu_{0}H_{\bot}=c_{\bot}%
^{2}/4D_{ex}\gamma$. We focus on the contribution of the $k_{\bot}$-- mode
(with $k_{\bot}=c_{\bot}/2D_{ex}$) to the transport coefficients~(\ref{generaltensor}) close to the touching field and expand  in
$\delta H=H-H_{\bot}$. As in Fig.~\ref{1DDIS}(b), for $k=k_{\bot}$ and $\delta
H\ll H_{\bot}$, the energies and group velocities of the upper and lower
magnon-polarons are approximately the same, i.e., $\Omega_{1k_{\bot}}%
\simeq\Omega_{2k_{\bot}}$ and $\partial_{k}\Omega_{1}|_{k=k_{\bot}}%
\simeq\partial_{k}\Omega_{2}|_{k=k_{\bot}}$. Eq. (\ref{31}) then reads
\begin{align}
W_{1k_{\bot}} &  =\frac{1}{2}\left[  1+\frac{\tilde{k}\delta H}{\sqrt
{1+(\tilde{k}\delta H)^{2}}}\right]\,,
\end{align}
with $\tilde{k}=\mu_{0}\gamma/(4 S_{\bot}k_{\bot})^{1/2}$. The scattering times
(\ref{tauik}) can be approximated as
\begin{align}
\tau_{1,2k_{\bot}}\sim\frac{\partial_{k}\Omega_{1,2k}|_{k=k_{\bot}}%
}{|v_{\mathrm{ph}}|^{2}}\frac{1}{(1-W_{1,2k_{\bot}})+\eta W_{1,2k_{\bot}}}\,.
\end{align}
Hence
\begin{align}
L_{xx}^{nm} &  \sim\frac{\beta}{L^{2}|v_{\mathrm{ph}}|^{2}}(\partial_{k}%
\Omega_{1k})^{3}\frac{e^{\beta\hbar\Omega_{1k}}}{(e^{\beta\hbar\Omega_{1k}%
}-1)^{2}}(\hbar\Omega_{1k})^{n}\bigg|_{\substack{k=k_{\bot},\\H=H_{\bot}%
}}\nonumber\\
&  \times y_{m}(\delta H)\,,\label{onemode}%
\end{align}
where
\[
y_{0}(\delta H)=\frac{4\left[ 1 + (\tilde{k} \delta H)^{2} \right](1+\eta)}{1+\eta \left[ 2 + 4 (\tilde{k} \delta H)^{2} + \eta \right]}\,,
\]
and
\[
y_{1}(\delta H)=\frac{2 \left[ 1 + 2 (\tilde{k} \delta H)^{2} + \eta \right]}{1+\eta \left[ 2 + 4 (\tilde{k} \delta H)^{2} + \eta \right]}\,.
\]
The indices $n$ and $m$ correspond to those in Eq.~(\ref{generaltensor}). Both
$y_{0}(\delta H)$ and $y_{1}(\delta H)$ have a single extremum at
$H=H_{\bot}$, i.e.,
\begin{align}
y_{0}^{^{\prime}}(\delta H)|_{\delta H=0} &  =y_{1}^{^{\prime}}(\delta
H)|_{\delta H=0}=0\,,\label{33}\\
y_{0}^{^{\prime\prime}}(\delta H)|_{\delta H=0} &  \propto(1-\eta
)^{2}\,,\label{34}\\
y_{1}^{^{\prime\prime}}(\delta H)|_{\delta H=0} &  \propto(1-\eta
)\,.\label{35}%
\end{align}
Eqs.~(\ref{33}) and (\ref{34}) prove that $y_{0}$ has a minimum at $H=H_{\bot
}$ for $\eta\neq1$, while for $\eta=1$ it is a constant. This explains our
numerical results for the heat conductivity $\kappa_{xx}$, which is
unstructured for $\eta=1$ and always display dips for both $\eta<1$ and
$\eta>1$ (see Fig.~\ref{fig6}(a)). According to Eqs.~(\ref{33}) and (\ref{35})
the function $y_{1}$ is also stationary at $H=H_{\bot}$, but it has a minimum only
for $\eta<1$, while an inflection point for $\eta=1$, and a maximum otherwise.
The resulting dependence on $\eta$ of Eq.~(\ref{onemode}) explains the spin
Seebeck coefficient $\zeta_{xx}$, the spin conductivity $\sigma_{xx}$ and
magnon heat conductivity $\kappa_{xx}^{(\mathrm{m})}$, in Figs.~\ref{3DBT}%
,~\ref{othercoefficients}(a) and \ref{othercoefficients}(b) respectively. As we have discussed in detail in the reporting of the numerical results, the anomalies can be understood physically in terms of the scattering time of the magnon-polaron. This scattering time is the sum of magnonic and phononic scattering times, so, depending on the value of $\eta$, the spin transport is enhanced ($\eta > 1$) or suppressed ($\eta < 1$) close to the touching point.

\subsection{Spin diffusion length}

Integrating the spin-projection of Eq.~(\ref{BOLTZ}) over momentum leads to
the spin conservation equation:
\begin{equation}
\dot{n}_{s}+\boldsymbol{\nabla}\cdot j_{s}=-g_{\mu}\mu\,,\label{EQZ26}%
\end{equation}
where
\begin{equation}
n_{s}=\int\frac{d^{3}\mathbf{k}}{(2\pi)^{3}}\sum_{i}f_{i{\mathbf{k}}%
}(\mathbf{r})\,,
\end{equation}
is the total magnon density (in units of $\hbar$), and
\begin{equation}
g_{\mu}=\beta\int\frac{d^{3}\mathbf{k}}{(2\pi)^{3}}\sum_{i}W_{i\mathbf{k}%
}\frac{1}{\tau_{i\mathbf{k}}^{\mathrm{nc}}}\frac{e^{\beta\hbar\Omega
_{i\mathbf{k}}}}{(e^{\beta\hbar\Omega_{i\mathbf{k}}}-1)^{2}}\,,\label{gnmu}%
\end{equation}
is the magnon relaxation rate, and we have introduced the relaxation time $\tau_{i\mathbf{k}}^{\mathrm{nc}}$. Elastic magnon-impurity scattering processes
discussed in the previous sections do not contribute to $\tau_{i\mathbf{k}%
}^{\mathrm{nc}}$. However, we parameterize the spin not-conserving
processes as
\begin{equation}
\frac{1}{\tau_{i\mathbf{k}}^{\mathrm{nc}}}=2\alpha\Omega_{\mathbf{k}%
i}\,,\label{newtau}%
\end{equation}
in terms of  the dimensionless Gilbert damping constant $\alpha$. In the non-equilibrium steady-state
Eq.~(\ref{EQZ26}) becomes
\begin{equation}
\boldsymbol{\nabla}^{2}\mu=\frac{1}{\lambda_{n}}\mu\,,
\end{equation}
in terms of the magnon diffusion length  $\lambda_{n}\equiv\sqrt{\sigma
_{xx}/g_{\mu}}$ that is plotted in Fig.~\ref{lambda}. At $10$ K and $50$~K,
the spin diffusion length decreases monotonously with the magnetic field for
$\eta=1$, in agreement with  observations at room temperature~\cite{ludo}. For
$\eta=100$ 
 ($\eta=0.01$) the spin diffusion length displays two peaks
(dips) at the critical fields $H_{\bot}$ and $H_{\parallel}$, which become
more pronounced when lowering the temperature. At $T=1$ K only the peak (dip)
at $H_{\bot}$ is visible for $\eta=100$ $(\eta=0.01)$. For $\eta=1$, the spin
diffusion length monotonically decreases with increasing magnetic field.  The
curve for $\eta=0.01$ behaves similar except for the dip at $H=H_{\bot}$. On
the other hand, for $\eta=100$, the spin diffusion length behaves very
differently showing strong enhancement at both low and high magnetic fields.
\begin{figure}[ptb]
\includegraphics[width=6.5cm]{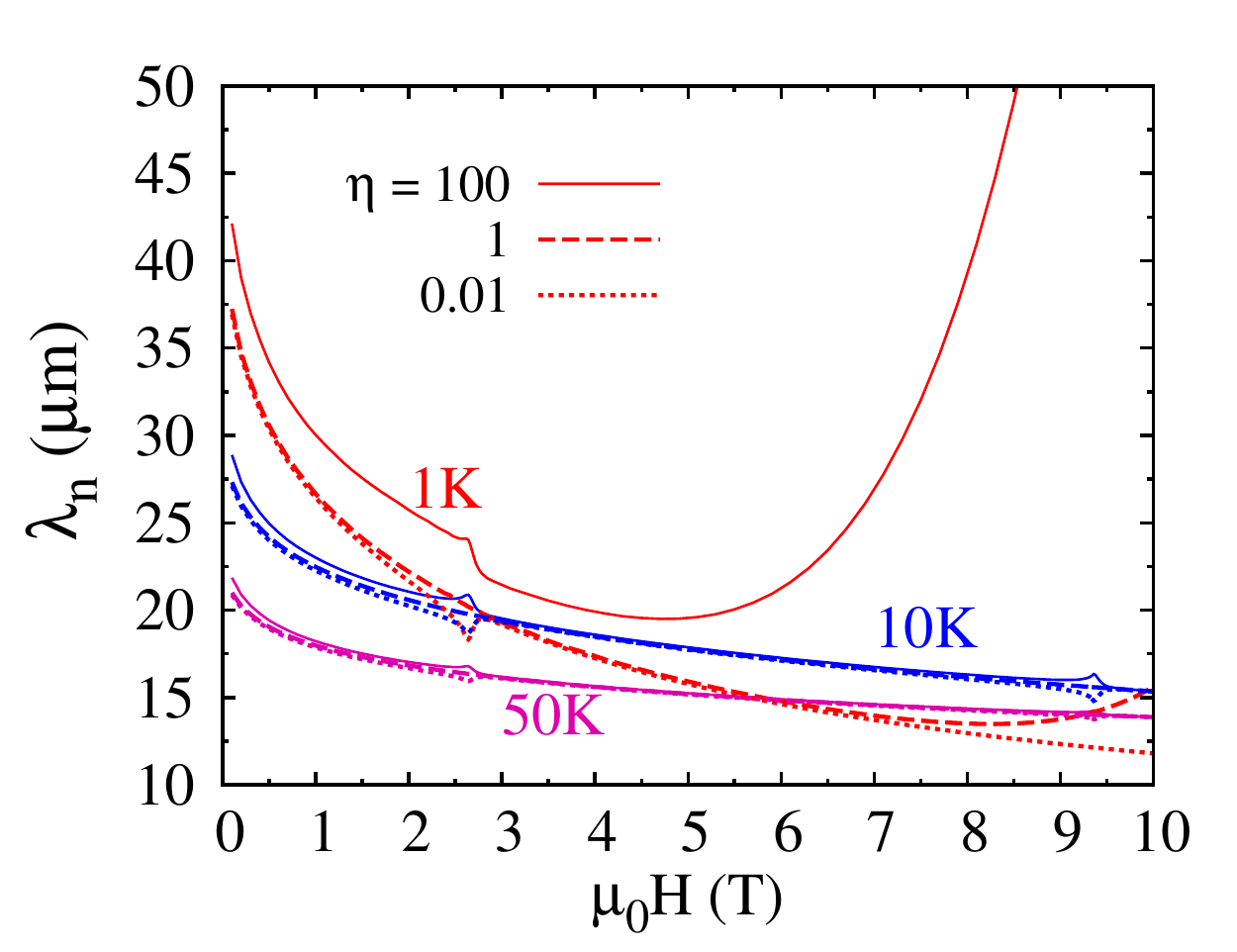} \caption{Magnetic field and
temperature dependence of the magnon diffusion length $\lambda_{n}$ for
different values of the scattering parameter $\eta$. }%
\label{lambda}%
\end{figure}This strong increase of the diffusion length (for constant Gilbert
damping) happens when
\begin{equation}
\frac{\sigma_{xx}(H_{1},\eta)}{\sigma_{xx}(H_{2},\eta)}>\frac{g_{\mu}(H_{1}%
)}{g_{\mu}(H_{2})}\,,\label{eqt36}%
\end{equation}
where $H_{1,2}$ are two given values of the applied magnetic field, with $H_{1}>H_{2}$.
 To understand the
dependence of the ratio $\sigma_{xx}(H_{1},\eta)/\sigma_{xx}(H_{2},\eta)$ on
$\eta$ and on the temperature, we recall that the main contribution to the
magnon spin conductivity $\sigma_{xx}$ arises from magnon-like branches. At
relatively high temperature, the magnon-like branches are sufficiently
populated to overcome the phonon contribution to the magnon spin conductivity
at all $\eta$. Indeed, Fig.~\ref{othercoefficients}(a) shows that, at
relatively high temperatures, the ratio $\sigma_{xx}(H_{1},\eta)/\sigma
_{xx}(H_{2},\eta)$ hardly depends on $\eta$. On the other hand, when the
temperature decreases below the magnon energy, the contribution of the
magnon-like branches are quickly frozen out by a magnetic field. The magnitude
of $\eta$ then becomes very relevant. On the other hand, while the right-hand
side of Eq.~(\ref{eqt36}) depends on temperature, it is not affected by $\eta
$. For $\eta<1$, the phonon mobility is smaller than the magnon one and hence
the phonons are short circuited by the magnons. For $\eta>1$, the phonons
prevail, leading to a higher ratio $\sigma_{xx}(H_{1},\eta)/\sigma_{xx}%
(H_{2},\eta)$ because the phonon dispersion is not affected by the magnetic
field. When $\eta\gg1$, the condition (\ref{eqt36}) is therefore satisfied. While in this regime the spin current is very small, it is perhaps an interesting limit for studying fluctuation and shot noise in the spin current~\cite{akash}.

\subsection{Comparison with experiments}

The spin Seebeck effect was measured in Pt$|$YIG$|$GGG structures in the
longitudinal configuration, i.e., by applying a  temperature difference normal
to the interfaces ($x$-direction) and subjecting the sample to a magnetic
field $\mathbf{H}\parallel\hat{\mathbf{z}}$~\cite{kikkawa}. The thermal bias induces a spin
current into the Pt layer that by the inverse Spin Hall effect (ISHE) leads to
the detected  transverse voltage $V$ over the contact, see Fig.~\ref{scheme}. The
bottom of the GGG substrate and the top of the Pt layer are in contact with
heat reservoirs at temperature $T_{L}$ and $T_{H},$ respectively. Disregarding
phonon (Kapitza) interface resistances, the phonon temperature gradient is
$\boldsymbol{\nabla} T=(T_{H}-T_{L})/L$, with $L$ being the thickness of the stack, and
average temperature $T=(T_{H}+T_{L})/2$. As discussed, we assume that the
magnon and phonon temperatures are the same and disregard the interface mixing
conductance. The measured voltage is then directly proportional to the bulk
spin Seebeck coefficient.

In the experimental temperature range of $3.5-50$\thinspace K the thermal
magnon, $\Lambda_{\mathrm{mag}}$, and phonon, $\Lambda_{\mathrm{ph},\mu}$,
wavelengths are of the order of $1-10$ \text{nm}. Even if the magnon and phonon thermal mean free paths have been estimated to be of the order of $\sim 100$ $\mu \text{m}$ at very low temperatures~\cite{Boona14}, here we assume  that the transport in the YIG
film of thickness $L\simeq4\,\mathrm{\mu}\text{m can  be treated
 semiclassically}$.  Note that scattering at the
 interfaces can make the transport diffusive even when the formal conditions for diffusive transport are not satisfied. The bulk spin Seebeck
coefficient is then well-described by Eq.~(\ref{generaltensor}) and
proportional to the observed voltage $V$. These assumptions are
encouraged by the good agreement for the observed and calculated peak
structures at $H_{\bot}$ and $H_{\parallel}$ with a single fitting parameter
$\eta=100$~\cite{kikkawa}. We may therefore conclude that the disorder potential scatters the
magnons more than the phonons and is therefore likely to be magnetic.

\section{Conclusion and Outlook}

We have established a framework which captures the effects of the
magnetoelastic interaction on the transport properties of magnetic insulators.
In particular, we show that the magnon-phonon coupling gives rise to peak-like
or dip-like structures in the field dependence of the spin and heat transport
coefficients, and of the spin diffusion length.

Our numerical evaluation reproduces the peaks in the observed low temperature
longitudinal spin Seebeck voltages of YIG%
$\vert$%
Pt layers as a function of magnetic field. We quantitatively explain the
temperature-dependent behavior of these anomalies in terms of hybrid
magnon-phonon excitations (\textquotedblleft magnon-polarons\textquotedblright%
). The peaks occur at magnetic fields and wave numbers at which the phonon
dispersion curves are tangents to the magnon dispersion, i.e., when magnon and
phonon energies as well as group velocities become the same. Under these
conditions the effects of the magnetoelastic interaction are
maximized. The
computed angle dependence shows a robustness of the anomalies with respect to
rotations of the magnetization relative to the temperature gradient. The
agreement between the theory and the experimental results confirms that
elastic magnon(phonon) impurity-scattering is the main relaxation channel that
limits the low temperature transport in YIG. Our theory contains one
adjustable parameter that is fitted to the large set of experimental data,
consistently finding a much better acoustic than magnetic quality of the
samples. The spin Seebeck effect is therefore a unique analytical instrument
not only of magnetic, but also mechanical material properties. The predicted
effects of magnon-polaron effects on magnonic spin and heat conductivity call
for further experimental confirmation.

We believe that the presented results open new avenues in spin caloritronics.
We focused here on the low energy magnon dispersion of cubic YIG, which is
well represented by the magnetostatic exchange waves of a homogeneous
ferromagnet~\cite{Barker2016}. However, the theoretical framework can be
easily extended to include anisotropies as well as ferri- or antiferromagnetic
order. The magnetoelastic coupling in YIG is relatively small and the
conspicuous magnon-polaron effects can be destroyed easily. However, in
materials with large magnon-phonon couplings these effects should survive in
the presence of larger magnetization broadening as well as higher temperatures.

\section{Acknowledgements}

This work was supported by the Stichting voor Fundamenteel Onderzoek der Materie (FOM), the European Research Council (ERC), the DFG Priority Programme 1538 \textquotedblleft Spin-Caloric Transport", Grant-in-Aid for Scientific Research on Innovative Area \textquotedblleft Nano Spin Conversion Science\textquotedblright\ (Nos. JP26103005 and JP26103006), Grant-in-Aid for Scientific Research (A) (Nos. JP25247056 and JP15H02012) and (S) (No. JP25220910) from JSPS KAKENHI, Japan, PRESTO \textquotedblleft Phase Interfaces for Highly Efficient Energy Utilization" and ERATO \textquotedblleft Spin Quantum Rectification Project" from JST, Japan,  NEC Corporation, and The Noguchi Institute. It is part of the D-ITP consortium, a program of the Netherlands Organization for Scientific Research (NWO) that is funded by the Dutch Ministry of Education, Culture and Science (OCW). T.K. is supported by JSPS through a research fellowship for young scientists (No. JP15J08026).




\end{document}